\documentclass[fleqn,usenatbib]{mnras}

\usepackage{newtxtext,newtxmath}

\usepackage[dvipsnames]{xcolor}
\usepackage{xcolor}
\definecolor{greyish}{HTML}{47555C}
\definecolor{redd}{HTML}{8F0200}
\definecolor{green}{HTML}{006400}
\definecolor{orange}{HTML}{ff7b00}
\definecolor{red1}{HTML}{BF100D}

\usepackage[T1]{fontenc}

\DeclareRobustCommand{\VAN}[3]{#2}
\let\VANthebibliography\thebibliography
\def\thebibliography{\DeclareRobustCommand{\VAN}[3]{##3}\VANthebibliography}

\usepackage{graphicx}	
\usepackage{amsmath}	

\newcommand{\HI}{\mbox{\textsc{H\,i}~}}
\newcommand{\Amod}{\mbox{$A_{\text{mod}}$}}
\newcommand{\Aflux}{\mbox{$A_{\text{flux}}$}}
\newcommand{\simba}{\textsc{simba}}
\newcommand{\martini}{\textsc{martini}}
\newcommand{\caesar}{\textsc{caesar}}

\title[\HI asymmetries in \textsc{simba} galaxies]{\HI asymmetries in spatially resolved \textsc{simba} galaxies}

\author[N. Hank et al.]{Nadine A. N. Hank,$^{1}$\thanks{E-mail: nhank@astro.rug.nl}
Marc A. W. Verheijen,$^{1}$
Sarah-L. Blyth,$^{2}$
Romeel Dav{\'e},$^{3,4}$
Kyle A. Oman,$^{5,6}$ \newauthor
Nathan Deg$^{7}$ and 
Marcin Glowacki$^{3,8,9}$
\\
$^{1}$Kapteyn Astronomical Institute, University of Groningen, Landleven 12, 9747 AD Groningen, the Netherlands\\
$^{2}$Department of Astronomy, University of Cape Town, Private Bag X3, Rondebosch 7701, South Africa\\
$^{3}$Institute for Astronomy, University of Edinburgh, Royal Observatory, Edinburgh EH9 3HJ, UK\\
$^{4}$Department of Physics and Astronomy, University of the Western Cape, Bellville, Cape Town 7535, South Africa\\
$^{5}$Institute for Computational Cosmology, Physics Department, Durham University, South Road, Durham DH1 3LE, UK\\
$^{6}$Centre for Extragalactic Astronomy, Physics Department, Durham University, South Road, Durham DH1 3LE, UK\\
$^{7}$Department of Physics, Engineering Physics, and Astronomy, Queen's University, Kingston, ON, K7L 3N6, Canada\\ 
$^{8}$International Centre for Radio Astronomy Research, Curtin University, Bentley, WA 6102, Australia\\ 
$^{9}$Inter-University Institute for Data Intensive Astronomy, Department of Astronomy, University of Cape Town, Cape Town, South Africa\\ 
}

\date{Accepted XXX. Received YYY; in original form ZZZ}

\pubyear{2025}

\begin{document}
\label{firstpage}
\pagerange{\pageref{firstpage}--\pageref{lastpage}}
\maketitle

\begin{abstract}
We present a study of the neutral atomic hydrogen (\HI\!) content of spatially resolved, low redshift galaxies in the \simba\ cosmological simulations. We create synthetic \HI data cubes designed to match observations from the Apertif Medium-Deep \HI imaging survey, and follow an observational approach to derive the \HI size-mass relation. The \HI size-mass relation for \simba\ is in broad agreement with the observed relation to within 0.1~dex, but \simba\ galaxies are slightly smaller than expected at fixed \HI mass. We quantify the \HI spectral ($\Aflux$) and morphological ($\Amod$) asymmetries of the galaxies and motivate standardising the relative spatial resolution when comparing $\Amod$ values in a sample that spans several orders of magnitude in \HI mass. Galaxies are classified into three categories (isolated, interacted, or merged) based on their dynamical histories over the preceding $\sim$2~Gyr to contextualise disturbances in their \HI reservoirs. We determine that the interacted and merged categories have higher mean asymmetries than the isolated category, with a larger separation between the categories' $\Amod$ distributions than between their $\Aflux$ distributions. For the interacted and merged categories, we find an inverse correlation between baryonic mass and $\Amod$ that is not observed between baryonic mass and $\Aflux$. These results, coupled with the weak correlation found between $\Aflux$ and $\Amod$, highlight the limitations of only using $\Aflux$ to infer the \HI distributions of spatially unresolved \HI detections.
\end{abstract}



\section{Introduction}

It has long been observed that peculiar and disturbed morphologies are prevalent in the stellar \citep[e.g.][]{Arp-66} and/or gas \citep[e.g.][]{Baldwin-80,Hibbard-01} distributions of galaxies. These asymmetries can arise as observational signatures of the myriad physical processes galaxies undergo over cosmic time, which also drive changes in other galaxy properties (e.g. colour, stellar mass, gas content, star formation rate (SFR), etc.). A number of these processes are associated with different environments and can have a significant impact on a galaxy's neutral atomic hydrogen (\HI\!) gas content, since it is susceptible to both gravitational and hydrodynamic forces. In low density environments, gas accretion from the cosmic web \citep[e.g.][]{Bournaud-05,Sancisi-08} and accretion of low-mass companion galaxies \citep{DiTeodoro-14} can replenish the gas in a galaxy, while fly-by interactions \citep{Mapelli-08} can remove gas from the outskirts of a galaxy. In high density environments, processes such as ram-pressure stripping \citep{Gunn-72}, viscous stripping \citep{Nulsen-82} and galaxy harassment \citep{Moore-96} deplete a galaxy's gas reservoir. Thus, studying \HI asymmetries, which have also been observed in the integrated \HI spectra \citep{Peterson-74, Haynes-98,Espada-11} and the \HI kinematics \citep{Swaters-99, Noordermeer-01, Chemin-06,Eymeren-11-1} of galaxies, could enhance our understanding of the evolutionary processes that these galaxies are undergoing. 

One of the first studies focusing on \HI asymmetries was undertaken by \cite{Peterson-74}, who examined the global \HI profiles of 23 galaxies with peculiar optical morphologies. They found a general correlation between the degree of asymmetry in the \HI profiles of these galaxies and the degree of asymmetry observed in their optical images, and speculated that the asymmetries were induced by tidal interactions or the presence of warps in the \HI discs. As \HI synthesis observations of nearby galaxies became available, \cite{Baldwin-80} were the first to highlight that large-scale morphological asymmetries are not just a feature observed in stellar discs after studying the spatial \HI distributions of $\sim$20 spiral galaxies. They reserved the term `lopsided' for any galaxy whose \HI distribution is disproportionately skewed towards one side of the galaxy, and proposed a kinematic model in which the observed lopsidedness is connected to off-centre elliptical orbits. Using data from a range of single dish \HI surveys, \cite{Richter-94} visually examined the \HI profiles of $\sim$1700 disc galaxies and found that at least 50~per~cent displayed strong or moderate asymmetric \HI profiles, suggesting that asymmetries are far more common than previously assumed. Since their sample consisted mainly of field galaxies to minimise the inclusion of interacting systems, the authors inferred that asymmetries are long lived, as opposed to frequently induced signatures. A similar result was found by \cite{Haynes-98}, who used the ratio of the integrated flux in two halves of the global \HI profiles to quantify asymmetry in 104 isolated galaxies, and found that $\sim$50~per~cent of their sample was considerably asymmetric. \cite{Matthews-98} found an even higher occurrence of asymmetry, albeit in a much smaller sample of 30 late-type spiral galaxies, with $\sim$77~per~cent displaying some degree of asymmetry in their \HI profiles. Although many of these studies were carried out using targeted galaxy samples, which tend to suffer from sample selection limitations, they remain crucial to our wider understanding of the ubiquity of \HI asymmetries. Nevertheless, untargeted \HI surveys are essential to get a more unbiased view of \HI asymmetries in the general galaxy population.\medskip 

The latest generation of untargeted \HI surveys on the Square Kilometre Array (SKA) precursor and pathfinder telescopes has been made possible by technological advances in the radio wavelength domain over the last two decades. Consequently, significant improvements in observing capabilities have been made. These surveys, which are arranged in a three-tiered `wedding cake' configuration in terms of their depth and sky area, will enable detailed studies exploring the evolution of \HI in galaxies using much larger samples than were previously available. Shallow surveys such as the Widefield ASKAP L-band Legacy All-sky Blind surveY \citep[WALLABY,][]{Koribalski-20} with the Australian SKA Pathfinder \citep[ASKAP,][]{Johnston-08,Hotan-21} and the Shallow Northern-sky Survey \citep[SNS,][]{Adams-19} with the APERture Tile In Focus \citep[Apertif,][]{Verheijen-08,VanCapellen-22} phased array feed system on the Westerbork Radio Synthesis Telescope (WSRT) form the bottom tier and will spatially resolve hundreds to thousands of nearby galaxies in \HI\!. The middle tier consists of the \HI emission project within the MeerKAT International GHz Tiered Extragalactic Exploration \citep[MIGHTEE-\textsc{H\,i},][]{Jarvis-16,Maddox-21} survey and the Medium-Deep \HI imaging Survey \citep[MDS,][]{Verheijen-09} with Apertif, which will observe local volumes to greater depth. The top tier of this arrangement is occupied by the Deep Investigation of Neutral Gas Origins \citep[DINGO,][]{Meyer-09} survey with ASKAP and the Looking At the Distant Universe with the MeerKAT Array \citep[LADUMA,][]{Blyth-16} survey, which will provide direct \HI detections out to $z < 1.4$. Other significant surveys aimed at directly detecting \HI in galaxies beyond $z = 0.15$ include the Blind Ultra-Deep \HI environmental Survey \citep[BUDHIES,][]{Jaffe-13} with the WSRT, which observed \HI in two $z \sim 0.2$ galaxy clusters, the HIGHz Arecibo survey \citep{Catinella-15}, which targeted massive, relatively isolated galaxies at similar redshifts to BUDHIES, and the COSMOS \HI Large Extragalactic Survey \citep[CHILES,][]{Fernandez-13}, which utilised the increased frequency coverage of the upgraded Karl G. Jansky Very Large Array and more than 1000 hours of observing time to push the redshift limit of \HI imaging studies out to $z=0.45$. Collectively, these surveys will enable much larger resolved galaxy samples at both low and high redshift to be studied for the first time. As a result, there has been renewed interest in \HI asymmetries in the lead up to these new datasets.

\cite{Espada-11} utilised the asymmetry index introduced in \cite{Haynes-98} to study the distribution of spectral asymmetry in a sample of extremely isolated disc galaxies from the Analysis of the interstellar Medium of Isolated GAlaxies \citep[AMIGA,][]{Montenegro-05} project, and found that the distribution could be described by a half-Gaussian with a mean of 1.0 and a standard deviation of $\sigma = 0.13$. They classified a galaxy as asymmetric if the measured value exceeded the $2\sigma$ threshold of their sample, and found that only 9~per~cent of their sample met this criterion. Subsequent studies using the same asymmetry metric have adopted the \cite{Espada-11} $2\sigma$ and $3\sigma$ thresholds to calculate the fractions of highly asymmetric global \HI profiles in their samples. For example, \cite{Scott-18} used the $3\sigma$ threshold and measured asymmetry fractions of 16~per~cent and 26~per~cent for galaxies with \HI detections in the Virgo and Abell 1367 clusters, respectively. \cite{Bok-19} compared samples of close galaxy pairs and isolated galaxies selected from the Arecibo Legacy Fast ALFA \citep[ALFALFA,][]{Haynes-18} survey, and found an enhanced asymmetry fraction in their close pair sample (27~per~cent) relative to their isolated sample (18~per~cent). \cite{Watts-20A} analysed the asymmetries of central and satellite galaxies from the extended GALEX Arecibo SDSS Survey \citep[xGASS,][]{Catinella-18} and found that satellite galaxies are generally more asymmetric than central galaxies. These results suggest that a correlation exists between the environment and the incidence of asymmetry. However, it remains unclear which astrophysical processes are primarily responsible for these asymmetries. 

A number of recent studies have introduced additional techniques to quantify \HI profile asymmetries \citep[e.g.][]{Deg-20,Reynolds-20A,Yu-20}. The use of these new methods in combination with the widely-used asymmetry index of \cite{Haynes-98} could offer valuable improvements in characterising the \HI profiles that will be available in large numbers. However, maximising the amount of information obtained from \HI profiles requires more knowledge of the underlying \HI distributions, since the shape of the \HI profile depends largely on the internal kinematics of the galaxy and to a lesser extent the spatial \HI distribution.

There are comparatively fewer quantitative asymmetry studies using \HI images of galaxies due to the lack of spatially resolved \HI observations available relative to \HI spectra. However, various methods that were developed for optical and near-infrared data have been adopted in \HI studies in recent years. One such method is the use of the normalised amplitude of the m = 1 Fourier component \citep[see][for a comprehensive review on this method]{Jog-09}. This technique was applied by \cite{Angiras-06, Angiras-07} to investigate morphological \HI asymmetries in the Eridanus group and the Ursa Major volume. They measured asymmetric fractions of $\sim$17$-$27~per~cent and found that the average asymmetry values measured in the outer regions of the \HI in the Eridanus sample was greater than the Ursa Major sample. Subsequently, \cite{Eymeren-11-2} used this method on a galaxy sample of from the Westerbork \HI survey of Irregular and SPiral galaxies \citep[WHISP,][]{Hulst-01} galaxies and found a continued increase in morphological asymmetry with radius for 20~per~cent of galaxies in their sample. 

Another method is the use of the concentration, asymmetry, smoothness, Gini and M$_{20}$ (CASGM) non-parametric measurements \citep{Conselice-3,Abraham-03,Lotz-04}. \cite{Holwerda-1,Holwerda-2} were the first to apply these measurements to \HI observations from The \HI Nearby Galaxy Survey \citep[THINGS,][]{Walter-08} and WHISP, and found that the asymmetry parameter was the most promising at identifying interacting galaxies in their sample. Following from this \cite{Giese-16} used simulated galaxies in order to better understand the observational effects on these non-parametric measurements. They observed that asymmetry was the ideal parameter to identify galaxies with lopsided \HI distributions, and determined a set of reliable resolution and inclination limits for the use of these parameters in upcoming large \HI surveys. Recent studies have also explored the use of these non-parametric measurements with machine learning techniques to identify morphologically disturbed galaxies in the WALLABY Pilot Survey \citep[e.g.][]{Holwerda-23,Holwerda-25}.

\cite{Wang-13} studied 23 nearby galaxies with high \HI mass fractions and compared them with a control sample of 19 galaxies. They found little difference in the overall asymmetry distributions of the populations, but noted that the outermost regions of the \HI\!-rich galaxies tended to be more asymmetric. \cite{Lelli-14} developed a modified outer asymmetry parameter that is more sensitive to low surface brightness regions on the outskirts of galaxies. They found that this modified asymmetry parameter was able to distinguish between samples of starburst dwarf galaxies and irregular galaxies.

More recently, \cite{Reynolds-20A} investigated the spectral, morphological and kinematic \HI asymmetries in $\sim$140 galaxies from three different surveys. They observed a slight trend of increasing incidence of asymmetry with increasing environment density, but found no correlation with \HI mass. They also found weak to moderate correlations between various asymmetry parameters, suggesting that the spectral asymmetry parameters are sufficient for spatially unresolved data, when the other parameters cannot be used. However, they concluded that a larger sample was required to confirm these trends. \medskip

In addition to these observational studies, a number of recent studies have used large-scale cosmological hydrodynamical simulations to investigate \HI asymmetries. These simulations offer significantly larger samples of spatially resolved galaxies, spanning a range of properties and cosmic environments, than are currently available with \HI surveys. Furthermore, they have the added benefit of providing detailed information about the galaxies' environments, their merger histories, and their progenitors at earlier cosmic epochs. Thus, simulations can offer important insights into the underlying astrophysical processes driving asymmetries.

\cite{Watts-20B} utilised the IllustrisTNG simulations \citep{Nelson-18} to examine the influence of environment on the profile asymmetries of $\sim$10500 galaxies, using halo mass as a proxy for environment. They observed a higher incidence of profile asymmetry among the satellite galaxy population than the central galaxy population, and found that this is driven by satellite galaxies residing in more massive haloes analogous to large galaxy groups. Using $\sim$3500 galaxies from the EAGLE simulations \citep{Schaye-15}, \cite{Manuwal-22} also found that the \HI profiles of satellite galaxies are generally more asymmetric than those of central galaxies, and determined that ram-pressure and tidal stripping are the strongest drivers of satellite asymmetry. \cite{Glowacki-22} used the \simba\ simulations \citep{Dave-19} to investigate trends between profile asymmetries and galaxy properties for $\sim$4000 galaxies. They found that the strongest correlation is between \HI mass and profile asymmetry, noting that the occurrence of profile asymmetries decreases with increasing \HI mass. 

\cite{Bilimogga-22} created mock \HI observations of $\sim$190 galaxies from the EAGLE simulations to investigate the combined effect of observational effects on spectral and morphological \HI asymmetries. They established various limits required for the resolution, \HI column density sensitivity and signal-to-noise of an observation to ensure robust asymmetry measurements. They also found no correlation between the spectral and morphological \HI asymmetries in their sample, seemingly in contradiction with the findings of \cite{Reynolds-20A}. Given that these studies both utilised relatively small ($N < 200$) samples, these conflicting results illustrate that a much larger sample is required to assess whether a correlation exists. \medskip

While various studies have recognised cosmological simulations as a useful tool to investigate \HI asymmetries and their limitations, research has yet to capitalize on the quantity of galaxies with spatially resolved \HI that is available with cosmological simulations. In this paper, we examine the \HI spectral and morphological asymmetries of $\sim$1100 spatially resolved, low redshift galaxies from the \simba\ cosmological simulations to understand and contextualise the state of disturbances in the \HI reservoirs of galaxies. Here, we consider the dynamical histories of galaxies over the preceding $\sim$2 Gyr, with a focus on investigating whether galaxy-galaxy interactions and mergers are the physical drivers of the \HI asymmetries observed in our sample at $z \sim 0$. We emphasise that our aim is not to determine the visibility time-scales over which these \HI asymmetries persist or to estimate merger rates using \HI data, but to compare and contrast the distributions of \HI spectral and morphological asymmetries of galaxies with different dynamical histories. In addition, we explore possible correlations between \HI asymmetry and baryonic mass in our simulated sample. 

\simba, which was calibrated to match the galaxy stellar mass function at $z=0$, successfully reproduces a number of galaxy observations at low redshifts \citep[e.g.][]{Dave-19,Li-19,Thomas-19, Appleby-20, Glowacki-20, Cui-21}. Most importantly for this work, \simba\ has been shown to reproduce the atomic and molecular gas contents of galaxies and shows good agreement with the \HI mass function at $z=0$ \citep{Dave-20}. We use the \martini\ package \citep{Oman-24} to generate synthetic \HI data cubes designed to match observations from the MDS. In doing so, we aim to enable a direct comparison between the \HI content of our simulated sample and that of the observed galaxy populations from the MDS, which will be explored in a future study. 

This paper is structured as follows. In Section~\ref{sec:Data}, we describe the key characteristics of the \simba\ simulations, the galaxy sample selection, the creation of mock \HI data cubes and the comparison of our sample to the observed \HI size-mass relation. Section~\ref{sec:Method} details the parameters used in this work to quantify spectral and morphological asymmetries, and our justification for standardising the relative spatial resolution across our sample when measuring and comparing morphological asymmetries. In Section~\ref{sec:DriversofAsymm} we present our main results, and compare our findings with existing studies in Section~\ref{sec:Discussion}. Finally, our conclusions are presented in Section~\ref{sec:Conclusions}. Throughout this paper, we use $H_0=68$~km~s$^{-1}$~Mpc$^{-1}$. 

\section{Data}\label{sec:Data} 

\subsection{The \textsc{simba} simulations }\label{sec:simba} 

The \simba\ simulations \citep{Dave-19} consist of a suite of cosmological hydrodynamical simulations of galaxy formation. These simulations solve for the co-evolution of galaxies, black holes, and intergalactic gas using a branched version of \textsc{Gizmo} \citep{Hopkins-15}, a gravity + hydrodynamics solver based on \textsc{Gadget-2} \citep{Springel-05}, in its meshless finite mass variant. In this study, we use the fiducial simulation (hereafter \mbox{\simba-100}), which evolves a representative (100$h^{-1}$~Mpc)$^3$ comoving volume initialised with $1024^3$ dark matter particles and $1024^3$ gas elements down to $z=0$. The resulting mass resolution is $9.6\times 10^{7}~\text{M}_{\odot}$ for dark matter particles and $1.82\times 10^{7}~\text{M}_{\odot}$ for gas elements. We supplement this with the high-resolution volume (hereafter \mbox{\simba-25}), which has identical input physics to \mbox{\simba-100} but has $8\times$ the mass resolution ($1.2\times 10^{7}~\text{M}_{\odot}$ for dark matter particles and $2.28\times 10^{6}~\text{M}_{\odot}$ for gas elements). This enables us to include lower mass galaxies in our sample. We refer the reader to \cite{Dave-19} for a comprehensive description of the simulations, and focus here on the key aspects of \simba\ and those most relevant to this study.

\simba\ follows a $\Lambda$-cold dark matter cosmology with parameters based on the \cite{Planck-16} results: $\Omega_{\text{m}}=0.3$, $\Omega_{\Lambda} =0.7$, $\Omega_{\text{b}}=0.048$, $H_0=68$~km~s$^{-1}$~Mpc$^{-1}$, $\sigma_8=0.82$, and $n_s=0.97$. The initial conditions are generated at a starting redshift of $z = 249$ using the \textsc{Music} code \citep{Hahn-11}. All simulation runs output 151 snapshots from $z =20 \rightarrow 0$, with a snapshot spacing of $\sim$230~Myr between $z = 0$ and $z=0.1$.

As the next generation of the \textsc{Mufasa} simulations \citep{Dave-16}, \simba\ builds on the work of its predecessor by incorporating a unique two-mode accretion model for black holes and implementing various improvements to the sub-grid prescriptions for star formation and feedback from active galactic nuclei (AGN). For cold gas ($T < 10^5$~K), black hole growth is implemented via a torque-limited accretion model \citep{Hopkins-11,Angles-17}, where instabilities in the cold gaseous disc give rise to gravitational torques which drive gas inflows, while Bondi accretion \citep{Bondi-52} is adopted for hot gas ($T > 10^5$~K). The accretion energy drives feedback that results in the quenching of galaxies, with a kinetic subgrid model for black hole feedback along with X-ray energy feedback included in \simba.

Radiative cooling and photoionisation heating are incorporated using the \textsc{grackle}-3.1 library \citep{Smith-17}, which includes metal cooling and non-equilibrium evolution of primordial elements. The chemical enrichment model follows 11 elements (H, He and 9 metals), with enrichment tracked from Type Ia and Type II supernovae as well as asymptotic giant branch stars. A \cite{Chabrier-03} initial mass function is used to compute stellar evolution. 

The total cold ($T<10^{5}$~K) neutral (\HI+ H$_2$) gas fraction is calculated self-consistently during the simulation runs by accounting for self-shielded gas on-the-fly, in lieu of applying self-shielding in the post-processing stage, as was the case in \textsc{Mufasa} \citep{Dave-17}. This is done using the prescription described in \cite{Rahmati-13}, where the metagalactic ionizing flux strength is attenuated based on the gas density, and assuming a spatially uniform ionizing background as given in \cite{Haardt-12}. 

The H$_2$ mass fraction for each gas element is given by
\begin{equation}
    f_{\text{H}_2} = 1-0.75\dfrac{s}{1+0.25s},
\end{equation}

\noindent with 
\begin{equation}
    s = \dfrac{\ln(1+0.6\chi+0.1\chi^{2})}{0.0396Z\Sigma}, 
\end{equation}

\noindent where $Z$ is the metallicity (in solar units) of the gas element, $\chi$ is the radiation field strength as a function of metallicity \citep{Krumholz-11}, and $\Sigma$ is the local column density (in M$_{\odot}$~pc$^{-2}$), which is estimated using the Sobolev approximation and modified to account for sub-resolution clumping \citep[see full discussion in][]{Dave-16}. Thereafter, the \HI mass fraction per gas element is determined by subtracting the H$_2$ fraction from the total cold neutral gas fraction.

Only gas elements above a Hydrogen number density threshold of $n_{\text{H}}>0.13$~H~atoms~cm$^{-3}$ and with $f_{\text{H}_2}>0$ are eligible for star formation. Gas above this density is considered the interstellar medium (ISM) and is artificially pressurised by imposing $T = 10^4 (n_{\text{H}}/0.13)^{1/3}$~K in order to resolve the Jeans mass \citep[see][]{Dave-16}. \simba's star formation prescription follows an H$_2$-based \cite{Schmidt-59} law and the SFR is given by 
\begin{equation}
    \text{SFR} = \dfrac{0.02\rho f_{\text{H}_2}}{\text{t}_{\text{dyn}}}, \label{eq:SFR}
\end{equation}

\noindent where $\rho$ is the gas density and t$_{\text{dyn}} = 1/\sqrt{G\rho}$ is the dynamical time. Star particles are spawned stochastically from eligible gas elements, where a gas element is converted into a star particle of the same mass and metallicty as the parent gas element. 

\subsection{Galaxy finding and sample selection}\label{sec:sample}

During the simulation runs, haloes are identified on-the-fly using a 3-D friends-of-friends (FoF) halo finder built into \textsc{Gizmo}, with a linking length equal to 0.2 times the mean inter-particle separation. Galaxies are identified separately using a 6-D FoF galaxy finder, with a spatial linking length of 0.0056 times the mean inter-particle separation (corresponding to twice the minimum softening length) and a velocity linking length equal to the local velocity dispersion. This is applied to all stars, black holes, and dense gas elements with $n_{\text{H}}>0.13$~H~atoms~cm$^{-3}$. The minimum stellar mass of a resolved galaxy is taken to be 32 star particle masses, above which galaxy properties can be reliably computed. This limit corresponds to $M_{\star,\text{min}} = 7.3\times 10^{7}$~M$_{\odot}$ for \mbox{\simba-25} and $M_{\star,\text{min}} = 5.8\times 10^{8}$~M$_{\odot}$ for \mbox{\simba-100}. 

\HI is assigned to galaxies in a separate step, instead of only using the gas elements assigned via the 6-D FoF galaxy finder. This decision is motivated by the presence of \HI in more diffuse, self-shielded gas which typically extends beyond the stellar radius of the galaxy \citep[see discussions in][]{Dave-19,Dave-20}. The \HI fraction is computed as described in Section \ref{sec:simba}, except all gas elements within a galaxy's halo with $n_{\text{H}}>0.001$~H~atoms~cm$^{-3}$ are considered. The \HI is then assigned via proximity to the galaxy to which these gas elements are most gravitationally bound. This allows for the \HI residing in gas outside of the star forming region of the galaxy to be included, while ensuring that no \HI is double-counted. 

Haloes and galaxies are cross-matched in post-processing using \caesar \footnote{\url{https://caesar.readthedocs.io/en/latest/}}, a python-based package that builds on the \texttt{yt} simulation analysis toolkit \citep{Turk-11}. \caesar\ outputs this information, in addition to pre-computed physical properties and particle lists of haloes and galaxies, in the form of a standalone HDF5 catalogue for each simulation snapshot. \medskip

For each volume, we use the snapshot at $z = 0.0167$ and its associated \caesar\ catalogue to select our sample. This snapshot is chosen because it is closest in redshift to Abell 262 \citep[$z=0.0163$,][]{Choque-21}, which is one of the galaxy clusters observed in the MDS that we aim to compare our mock sample with in a future study. We impose a lower stellar mass limit of $M_{\star} \geq 7.3\times 10^{7}$~M$_{\odot}$ to \mbox{\simba-25} and $M_{\star} \geq 5.8\times 10^{8}$~M$_{\odot}$ to \mbox{\simba-100}, which corresponds to the minimum stellar mass of a resolved galaxy in each volume. We also impose a lower \HI mass limit of $M_{\text{\textsc{H\,i}}} \geq 2.28\times 10^{8}$~M$_{\odot}$ to \mbox{\simba-25} and $M_{\text{\textsc{H\,i}}} \geq 1.82\times 10^{9}$~M$_{\odot}$ to \mbox{\simba-100}, to ensure that the \HI content of a galaxy is composed of at least 100 gas elements. We apply this \HI mass limit to the \HI mass that is confined to the dense ISM gas of a galaxy as opposed to the galaxy's total \HI mass provided by the \caesar\ catalogue. This decision is motivated by the method used to generate synthetic \HI data cubes in this work. \martini\ (described further in Section~\ref{sec:martini}) is used to extract gas elements in a spherical aperture that is centred on the minimum of a galaxy's gravitational potential when loading a galaxy's \HI content into a data cube. However, the method \simba\ uses to assign \HI to galaxies means that it is possible for a massive galaxy to have \HI in its surrounding environment that has been assigned to it, but no \HI located in the central, star forming region of the galaxy. This will result in an empty data cube if the galaxy's \HI content lies entirely outside \martini's extraction aperture. One solution would be to increase the aperture used, but this slows down the cube creation considerably, increases the likelihood of including \HI emission from neighbouring galaxies in the foreground/background, and does not guarantee that all of the galaxy's \HI content will be contained in the data cube. Instead, by imposing the \HI cut to the \HI mass in the central region of the galaxy, we ensure that a significant fraction of a galaxy's \HI is still associated with the galaxy itself. We also performed consistency checks to ensure that the extracted cubes were large enough to encompass the full \HI distributions of the galaxies (see Section~\ref{sec:sampleRefine}).

The specified stellar and \HI mass cuts are applied to each volume, resulting in an initial sample of 669 galaxies from \mbox{\simba-25} and 5914 galaxies from \mbox{\simba-100}. Due to the larger effective volume of \mbox{\simba-100} and limited computing resources, we reduce the sample size of this volume to match that of \mbox{\simba-25} by randomly selecting 669 galaxies from the 5914 galaxies such that the \HI mass distribution of the downsized sample reproduces the distribution of \mbox{\simba-100's} initial sample. Nontheless, our starting sample of 1338 galaxies (and the final 1157 galaxies after the sample refinement described in Section~\ref{sec:sampleRefine}) is $> 5\times$ larger than previous studies investigating \HI morphological asymmetries \citep[e.g.][]{Reynolds-20A,Bilimogga-22} and sufficient for the objectives of this study. Our procedure for generating mock \HI data cubes for the starting sample is described in the following section.

\subsection{Synthetic \textsc{H\hspace{.122em}i} data cubes}\label{sec:martini}

\martini\footnote{\url{https://github.com/kyleaoman/martini}, version 1.5 (git commit c455f17). Subsequent versions of \martini\ released after the analysis for this work was carried out include some minor changes that would slightly affect our measurements. However, we have checked a subset of galaxies with code versions 2.0.0 and 2.0.6 and find that the changes are minimal and do not impact our conclusions.} \citep{Martini-code,Oman-24} is a Python package developed for creating synthetic, spatially and kinematically resolved \HI data cubes from smooth particle hydrodynamical simulations. It includes instrument-specific submodules for the data cube, beam, noise, source and spectral model, which can be configured to reproduce mock interferometric \HI observations of galaxies. The methodology used to create \HI data cubes with \martini\ is described in \cite{Oman-19}. Here we provide a summary that is specific to our sample of galaxies. 

In \martini, we specify a fixed aperture with a radius of 100 kpc to extract the galaxy of interest and the region surrounding it from the simulation snapshot. The specific angular momentum vector of the target galaxy's \HI disc, $\vec{L}_{\text{\textsc{H\,i}}}$, is calculated using the central 1/3 of the \HI disc (by mass). The disc plane is then oriented in the y-z plane by aligning the x-axis with the direction of $\vec{L}_{\text{\textsc{H\,i}}}$. Each galaxy is given a randomly-selected azimuthal rotation (around $\vec{L}_{\text{\textsc{H\,i}}}$) and inclination angle with respect to the line of sight. For all galaxies in our sample, the kinematic position angle is set to 270\textdegree, measured counter-clockwise from North. 

Galaxies are placed in the Hubble flow at a distance of 71.86~Mpc, the adopted distance to Abell 262 \citep{Choque-21}, resulting in an angular scale of $\sim$348 parsec per arcsecond. Each data cube, which is centred on the minimum of the target galaxy's gravitational potential, is generated with $175 \times 175$ pixels and a pixel size of $5\arcsec \times5 \arcsec$. A velocity channel spacing of 7.86~km~s$^{-1}$ is used and 128 velocity channels are included to ensure that the full extent of the \HI emission of \HI massive galaxies is contained within the data cubes. The data cubes are convolved to a \mbox{$15\arcsec$-FWHM} circular Gaussian beam, which corresponds to $\sim$1.74~kpc per pixel at the distance of Abell 262. These velocity and angular resolutions are chosen to match the observational parameters of the MDS data cubes. The unit of flux in the data cubes is ${\text{Jy~beam}^{-1}}$. For the purposes of this study, no instrumental noise is added to the data cubes.

\subsection{Sample refinement}\label{sec:sampleRefine}

Because we extract all gas elements within 100~kpc, the \HI of a neighbouring galaxy will be included in the target galaxy's data cube if the galaxies are within 100 kpc and (128/2)~$\times$~7.86~km~s$^{-1}$ of each other. We use the OBJECTS task from \texttt{GIPSY}\footnote{\url{https://www.astro.rug.nl/~gipsy}} \citep{Hulst-92,Vogelaar-01} to remove the \HI emission from companion galaxies to prevent this extraneous \HI from contributing to the \HI mass and the asymmetry parameters measured for the galaxies in our sample. For each data cube, OBJECTS was set to identify all independent structures above a column density threshold of $1\times 10^{19}$~cm$^{-2}$ over 25~km~s$^{-1}$ and present in at least 5 channels. The largest 3D structure identified by OBJECTS that is closest to the centre of the data cube (spatially and spectrally) was used as a mask to separate the target galaxy from all other detected objects in the data cube. For each galaxy, the masked data cube was then used to measure the \HI mass using the formula

\begin{equation}
    \Bigg ( \dfrac{ M_{\text{\mbox{\textsc{H\,i}}}}}{\text{M}_{\odot}}  \Bigg ) \simeq \dfrac{2.35\times 10^{5}}{(1+z)} \Bigg ( \dfrac{D}{\text{Mpc}} \Bigg )^{2}  \Bigg ( \dfrac{S}{\text{Jy km s}^{-1}}  \Bigg ),
\end{equation}

\noindent where $z$ is the redshift of the galaxy (set to $z=0.0163$), $D$ is the proper distance (71.86~Mpc) and $S = \int S_{v} dv$ is the integrated \HI line flux density measured in the data cube \citep{Meyer-17}. 

\begin{figure}
    \includegraphics[width=\columnwidth]{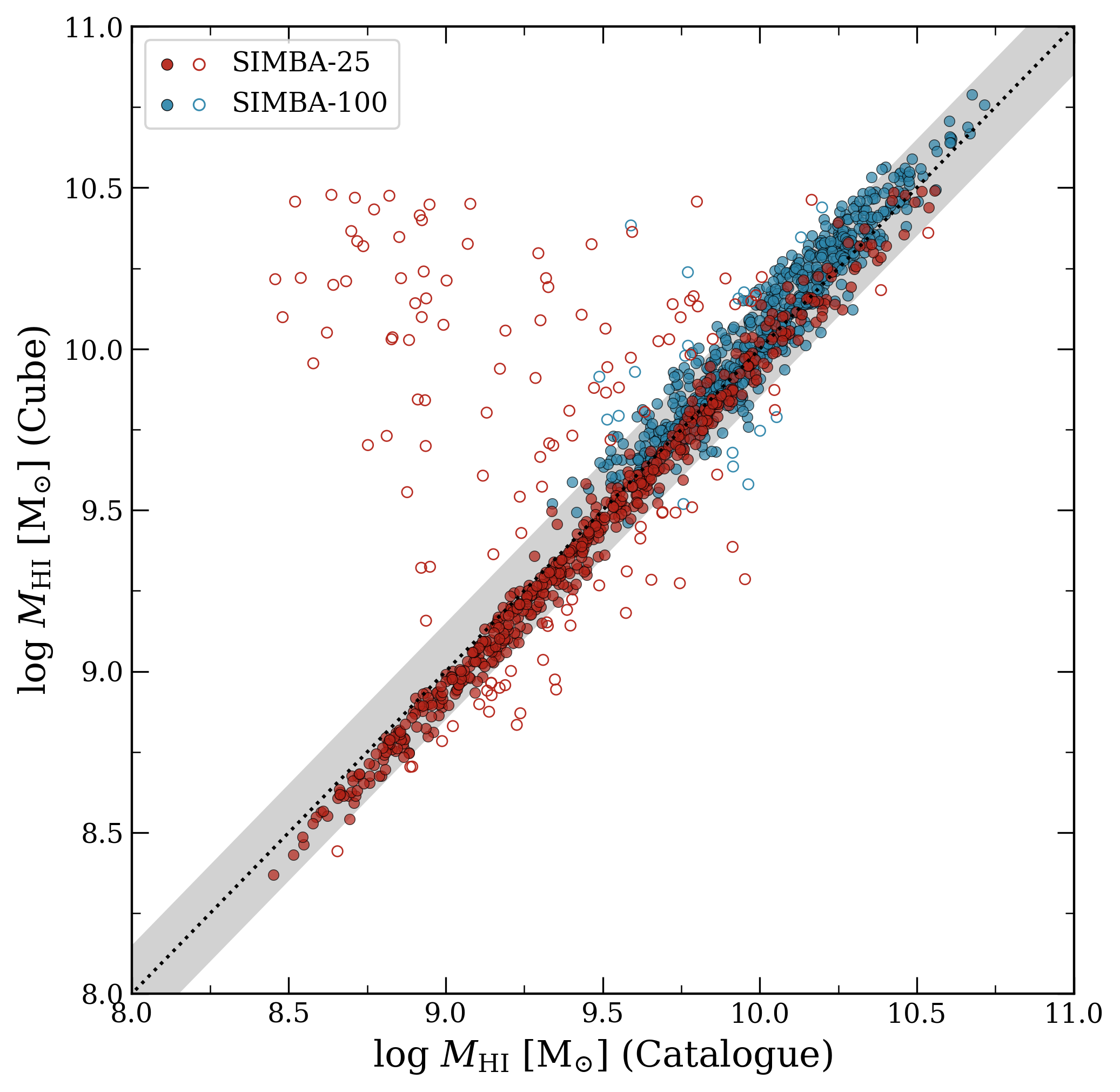}
    \caption{Comparison between the \HI mass measured from the masked mock data cube and the expected \HI mass from the \caesar\ catalogue for the \mbox{\simba-25} (red) and \mbox{\simba-100} (blue) galaxies. The black dotted line shows the 1:1 relation, with the grey shaded region representing a 0.15~dex buffer from the relation. Galaxies that deviate from the 1:1 relation by more than 3$\sigma_{\text{MAD}}$ are plotted as open circle markers and have been excluded from the samples.}
    \label{fig:HI-massComp}
\end{figure}

In Fig.~\ref{fig:HI-massComp} we compare the measured \HI masses from our masked data cubes to the expected \HI masses from the \caesar\ catalogues. For both volumes' samples, the majority of galaxies are scattered within $\sim$0.15 dex of the 1:1 relation, illustrating that \simba's approach to assigning \HI to galaxies yields reasonable \HI masses that are in agreement with the \HI masses measured from the mock data cubes. This also indicates that most of the \HI content of the galaxies is successfully recovered after masking the data cubes. However, a considerable number of galaxies, mostly from the \mbox{\simba-25} sample, have much higher ($\geq0.5$~dex) measured \HI masses than expected. Upon inspecting their cubes, we find that most of these outliers are satellite galaxies whose central galaxy has remained in the data cube due to connecting bridges of \HI between the galaxies. A higher column density threshold of $3.5\times 10^{19}$~cm$^{-2}$ was used to test if the target galaxies could be disconnected from their larger companions, but this threshold was either unsuccessful in separating the galaxies or resulted in considerable amounts of low column density \HI being masked out, causing galaxies to scatter downwards from the 1:1 relation. It is also worth noting that the \mbox{\simba-100} sample appears slightly offset upwards from the 1:1 relation by $\sim$0.1~dex, whereas the \mbox{\simba-25} sample lies slightly below the 1:1 relation. Although a higher column density threshold would shift the \mbox{\simba-100} galaxies closer to the 1:1 relation, this would result in the \mbox{\simba-25} sample shifting further away from the relation since the same threshold is used to ensure that both samples are treated consistently.

We opted to remove the significantly outlying galaxies from both samples for a number of reasons. First, these galaxies will result in artificially high \HI asymmetries that are inflated due to the presence of multiple sources in the data cubes and not the physical processes disturbing the \HI discs. Second, when investigating the dynamical histories of galaxies (see Section~\ref{sec:DriversofAsymm} for more details), we consider the fractional changes in a galaxy's baryonic mass (calculated using the stellar and \HI masses) over several snapshots up to a redshift of $z\sim 0.2$. However, we do not generate cubes at these snapshots and rely solely on the masses provided in the \caesar\ catalogues. By restricting our samples to galaxies with measured \HI masses that are in agreement with the catalogue \HI masses, we reduce the likelihood of under- or overestimating the \HI masses (and therefore the baryonic masses) at previous snapshots.

To identify outliers in each sample, we use the robust estimator $\sigma_{\mathrm{MAD}} = 1.4826\text{MAD}$, where MAD is the median absolute deviation, because it is less sensitive to the presence of outliers than the standard deviation around the mean. Galaxies that deviate from the 1:1 \HI mass relation by more than $3\sigma_{\text{MAD}}$ (shown as open circle markers in Fig.~\ref{fig:HI-massComp}) are excluded from the samples. We further remove galaxies that cannot be reliably traced in the snapshots used to determine their dynamical histories (see Section~\ref{sec:DriversofAsymm}). These cuts remove a total of 181 galaxies, resulting in the final 513 galaxies from \mbox{\simba-25} and 644 galaxies from \mbox{\simba-100} which are studied in the remainder of this work. The \HI total intensity (moment-0) maps and global \HI profiles are obtained from the masked cubes by integrating along the spectral axis and by summing the flux in each channel, respectively. 

\begin{figure*}
 \centering\includegraphics[width=\linewidth]{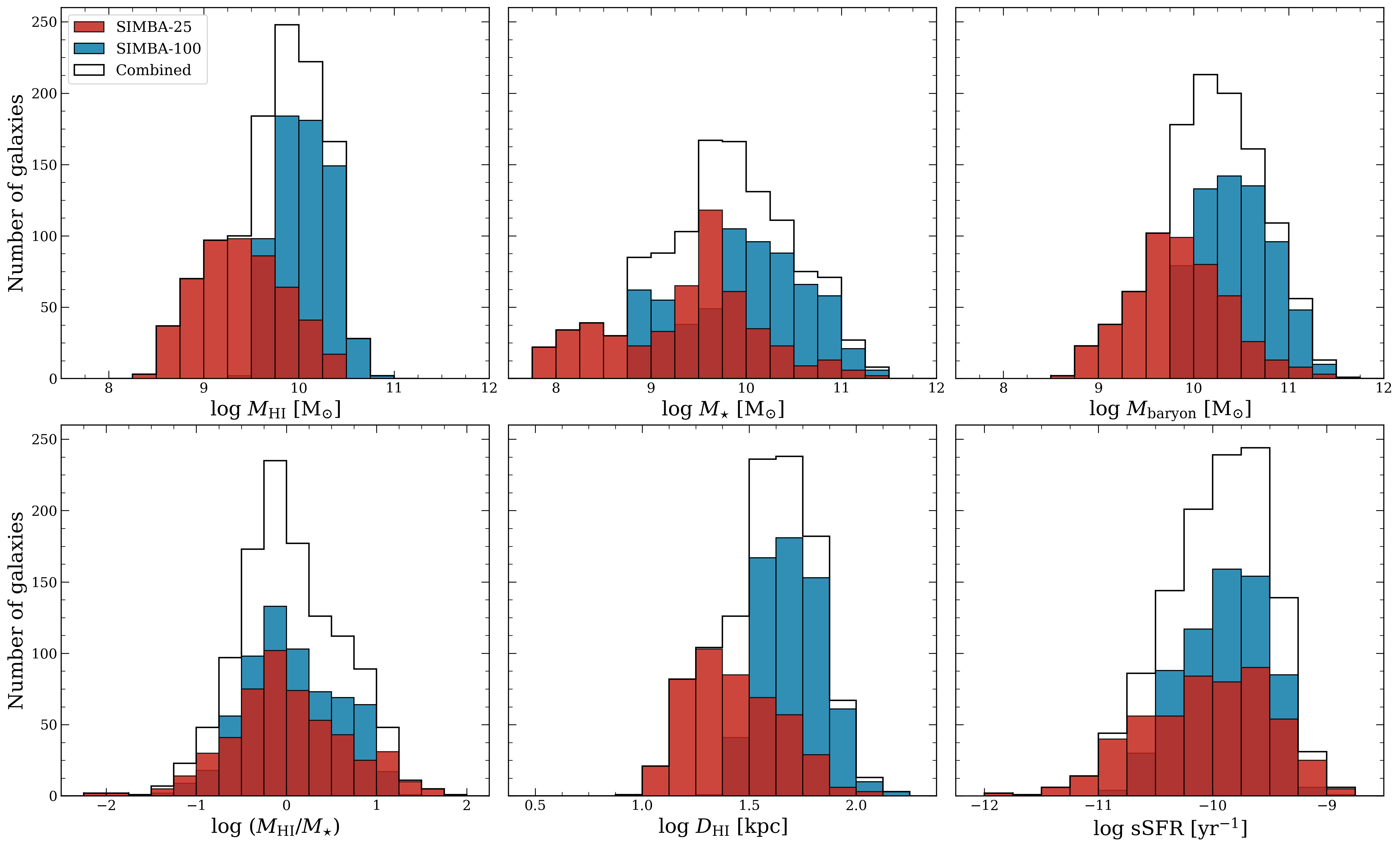}
 \caption{Sample properties for the final sample of \mbox{\simba-25} (red) and \mbox{\simba-100} (blue) galaxies. \textit{Top row:} Distributions of the measured \HI mass, stellar mass, and baryonic mass (calculated as $M_{\text{baryon}} = M_{\star} +1.33M_{\text{\textsc{H\,i}}}$). \textit{Bottom row:} Distributions of the \HI gas fraction, \HI diameter, and specific star formation rate (sSFR = SFR/$M_{\star}$). In all panels, the black step histogram represents the combined sample. The stellar masses and specific star formation rates are taken from the \caesar\ catalogue. All other properties related to the \HI content make use of the \HI from the masked cubes. The \HI diameter distributions only include galaxies that are resolved by $\geq 2$~beams across the minor axis (see Section~\ref{sec:DataVerification}).}
 \label{fig:FS_massDist}
\end{figure*}

Fig.~\ref{fig:FS_massDist} shows the distributions of the \HI mass, stellar mass, baryonic mass ($M_{\text{baryon}} = M_{\star} +1.33M_{\text{\textsc{H\,i}}}$, where the factor of 1.33 accounts for the presence of helium), \HI gas fraction ($M_{\text{\textsc{H\,i}}}/M_{\star}$), \HI diameter, and specific star formation rate (sSFR = SFR/$M_{\star}$) for the final galaxy samples. The stellar masses and SFRs are taken from the \caesar\ catalogues at $z=0.0167$. All properties relating to the \HI mass are obtained using the masked \HI data cubes.

\subsection{Data verification}\label{sec:DataVerification}

Before proceeding further, we quantify the spatial distributions of \HI in our samples using the \HI radius, $R_{\text{\mbox{\textsc{H\,i}}}}$, and assess \simba's ability to reproduce the well-studied \HI size-mass relation. Observationally, it has been shown that a tight relation exists between the total \HI mass of a galaxy and the diameter of its \HI disc \citep[e.g.][]{Broeils-97,Verheijen-01,Begum-08,Wang-16,Lelli-16,Naluminsa-21,Rajohnson-22}. The relation can be described by 

\begin{equation}
    \text{log } \Bigg ( \dfrac{ D_{\text{\mbox{\textsc{H\,i}}}}}{\text{kpc}}  \Bigg ) = m \text{ log } \Bigg ( \dfrac{ M_{\text{\mbox{\textsc{H\,i}}}}}{\text{M}_{\odot}}  \Bigg ) + c,
\end{equation}

\noindent where $D_{\text{\mbox{\textsc{H\,i}}}}$ is the diameter of the galaxy measured at an \HI surface mass density of 1~M$_{\odot}~$pc$^{-2}$. The \HI size-mass relation holds irrespective of morphological type and galaxy size, suggesting that the mean \HI surface density remains approximately constant and that galaxies likely evolve along the relation. 

Several studies utilising hydrodynamical simulations \citep{Bahe-16,Marinacci-17,Diemer-19,Stevens-19,Gensior-24} and semi-analytic models \citep{Obreschkow-09,Wang-14,Lutz-18} have found remarkable success in reproducing the observed relation with simulated data at low redshift, albeit with some variations in the scatter, slope, and intercept. Since \simba\ dynamically tracks \HI instead of applying a post-processing correction, and does not fine-tune any parameters to reproduce observed \HI scaling relations, the robustness of the \HI size-mass relation makes it an essential benchmark to evaluate how the subgrid prescriptions and feedback mechanisms implemented in \simba\ affect the \HI content of galaxies.

We follow an observational approach by making use of the moment-0 maps to derive azimuthally averaged radial \HI surface density profiles. This is done with the ELLINT task from \texttt{GIPSY}, which integrates the moment-0 map in concentric ellipses. The position and inclination angles of the ellipses are set to the values used when generating the data cubes in Section~\ref{sec:martini}. Each radial profile is deprojected to the face-on equivalent and then scaled by the total \HI mass of the galaxy. To get $R_{\text{\mbox{\textsc{H\,i}}}}$, we interpolate the profiles and measure the radius at which the \HI surface mass density ($\Sigma_{\text{\mbox{\textsc{H\,i}}}}$) is equal to 1~M$_{\odot}~$pc$^{-2}$ ($\sim \! 1.249\times 10^{20}$~cm$^{-2}$). It should be noted that this procedure is impacted by beam smearing. If the minor axis of a galaxy’s \HI disc is not sufficiently resolved, this will lead to the `smearing’ out of \HI emission and a seemingly rounder \HI disc. A galaxy's inclination angle is used to determine the ellipticity of the ellipses used to derive its radial profile, as well as the correction factor to deproject the profile to represent face-on values. If the outermost ellipse does not enclose the full extent of the rounder \HI disc due to beam smearing, this will lead to underestimating $R_{\text{\mbox{\textsc{H\,i}}}}$ because the total \HI mass of the galaxy is used to scale the profile to surface mass densities. Alternatively, if an edge-on galaxy has a warp in its \HI disc or extraplanar \HI that falls outside the outermost ellipse, $R_{\text{\mbox{\textsc{H\,i}}}}$ will also be underestimated for the same reason. Thus, to mitigate the effects of beam smearing, we restrict our analysis in this section to galaxies that are resolved by $\geq 2$ beams across the minor axis, leaving us with 456 galaxies from \mbox{\simba-25} and 617 galaxies from \mbox{\simba-100}.

\begin{figure}
	\includegraphics[width=\columnwidth]{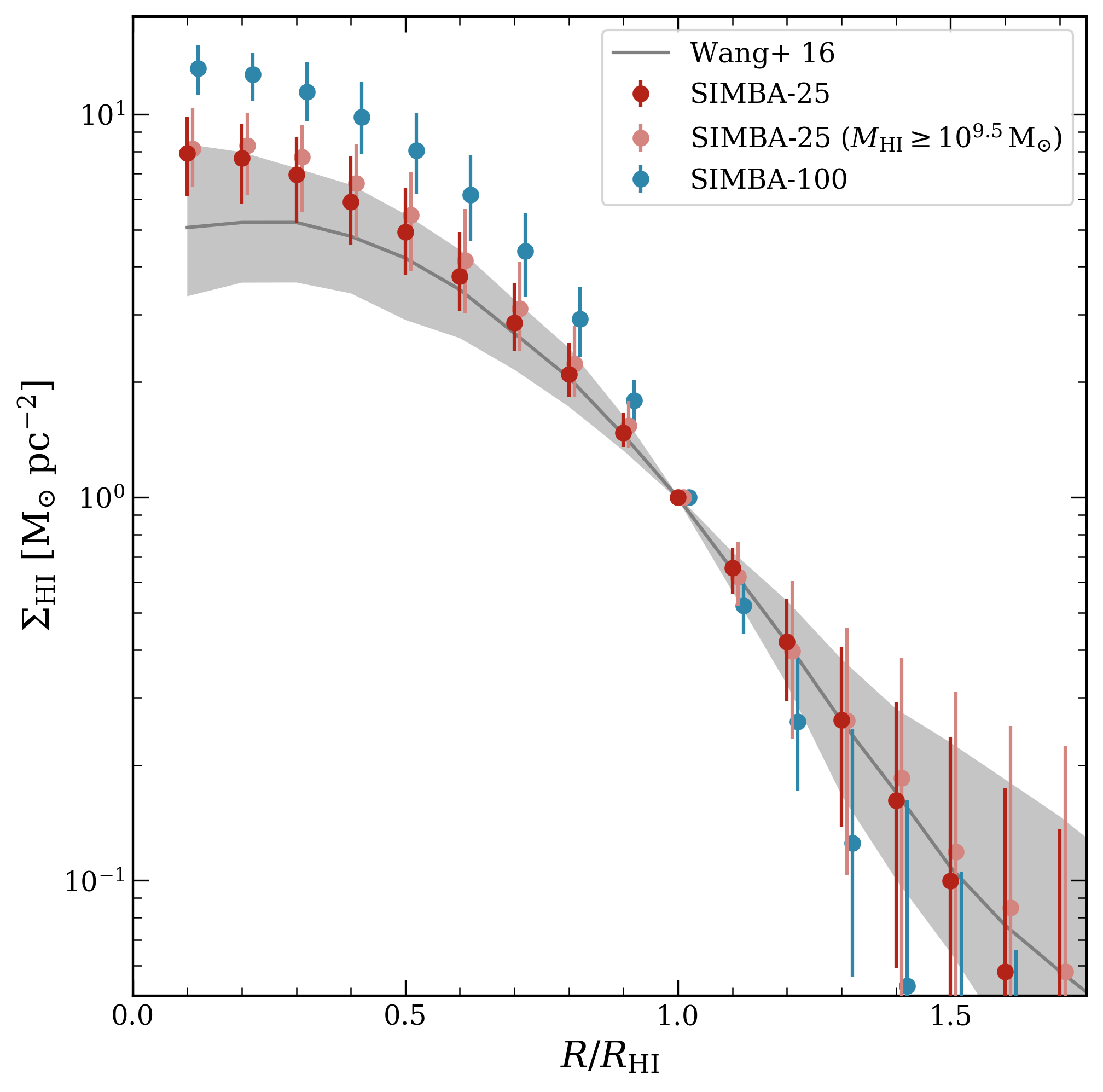}
    \caption{Median azimuthally averaged \HI surface density profiles for galaxies from \mbox{\simba-25} (red) and \mbox{\simba-100} (blue). The median radial profile of high-\HI mass galaxies ($M_{\text{\textsc{H\,i}}} \geq 10^{9.5}~\text{M}_{\odot}$) from \mbox{\simba-25} is shown in pink and has been slightly offset horizontally, along with the median profile of \simba-100, for easier viewing. The solid grey line shows the median radial profile for a combined sample of resolved galaxies from \protect\cite{Wang-16} (we refer the reader to their paper for further details on the \HI surveys used to compile the sample). The error bars and the shaded region indicate the interquartile range of the three samples.}
    \label{fig:medRadProf}
\end{figure}

Fig.~\ref{fig:medRadProf} shows the median radial profiles, normalised to $R_{\text{\mbox{\textsc{H\,i}}}}$, for the \simba-25 (red) and \mbox{\simba-100} (blue) samples. To investigate if any differences between the median profiles of the two samples is a result of their different \HI mass ranges, we also show the median radial profile of \mbox{\simba-25} galaxies with $M_{\text{\textsc{H\,i}}} \geq 10^{9.5}~\text{M}_{\odot}$, where the two samples overlap. For an observational comparison, we include the median radial profile of a subsample of galaxies from \cite{Wang-16} that are resolved by $\geq 3$ beams across the major axis (see their fig. 2 for the \HI surveys used to compile the sample). For all samples, the median values are computed in bins of $0.1\times R/R_{\text{\mbox{\textsc{H\,i}}}}$. The data for the high-\HI mass \mbox{\simba-25} subsample and \mbox{\simba-100} have been slightly offset horizontally for visualisation purposes. The error bars and shaded area indicate the interquartile range (i.e. the region between the 25$^{\text{th}}$ and 75$^{\text{th}}$ percentiles) of each sample.

Qualitatively, the median profiles of \mbox{\simba-25} and \mbox{\simba-100} galaxies broadly agree with the median profile from observed data in terms of their shapes, but there are apparent differences between the simulated and observed samples in the innermost regions of the \HI discs. \mbox{\simba-100} galaxies have systematically higher central \HI surface densities than observed in the local Universe, with a steeper decline in surface density beyond $R_{\text{\mbox{\textsc{H\,i}}}}$. Galaxies from \mbox{\simba-25} also display slightly higher central \HI surface densities in the inner regions ($R \lesssim 0.5 R_{\text{\mbox{\textsc{H\,i}}}}$), but the median profile falls within the overlap of the interquartile ranges of the radial profiles from \mbox{\simba-25} and the observational sample, and largely agrees with observations at larger radii ($R \gtrsim R_{\text{\mbox{\textsc{H\,i}}}}$). While the median radial profile of the high-\HI mass \mbox{\simba-25} subsample has marginally higher \HI surface densities than that of the full \mbox{\simba-25} sample, the profiles are generally consistent. However, despite having similar \HI mass ranges, the median profiles of the high-\HI mass \mbox{\simba-25} subsample and \mbox{\simba-100} are still offset from each other, indicating that the difference in the \HI mass ranges of \mbox{\simba-25} and \mbox{\simba-100} is not the primary cause of the offsets between the two simulated samples.

The observed differences between the median profiles of the \mbox{\simba-25} and \mbox{\simba-100} samples are possibly due to the difference in the numerical resolutions of the two volumes. As mentioned, \simba\ incorporates a self-shielding approximation from \cite{Rahmati-13} that depends on the local gas density in order to compute the \HI fractions. It is likely that the higher particle numbers for galaxies in \mbox{\simba-25} relative to galaxies of similar \HI mass in \mbox{\simba-100} results in more self-shielding in the former volume, and thus more \HI in the outskirts of the \HI discs. 

\begin{figure}
	\includegraphics[width=\columnwidth]{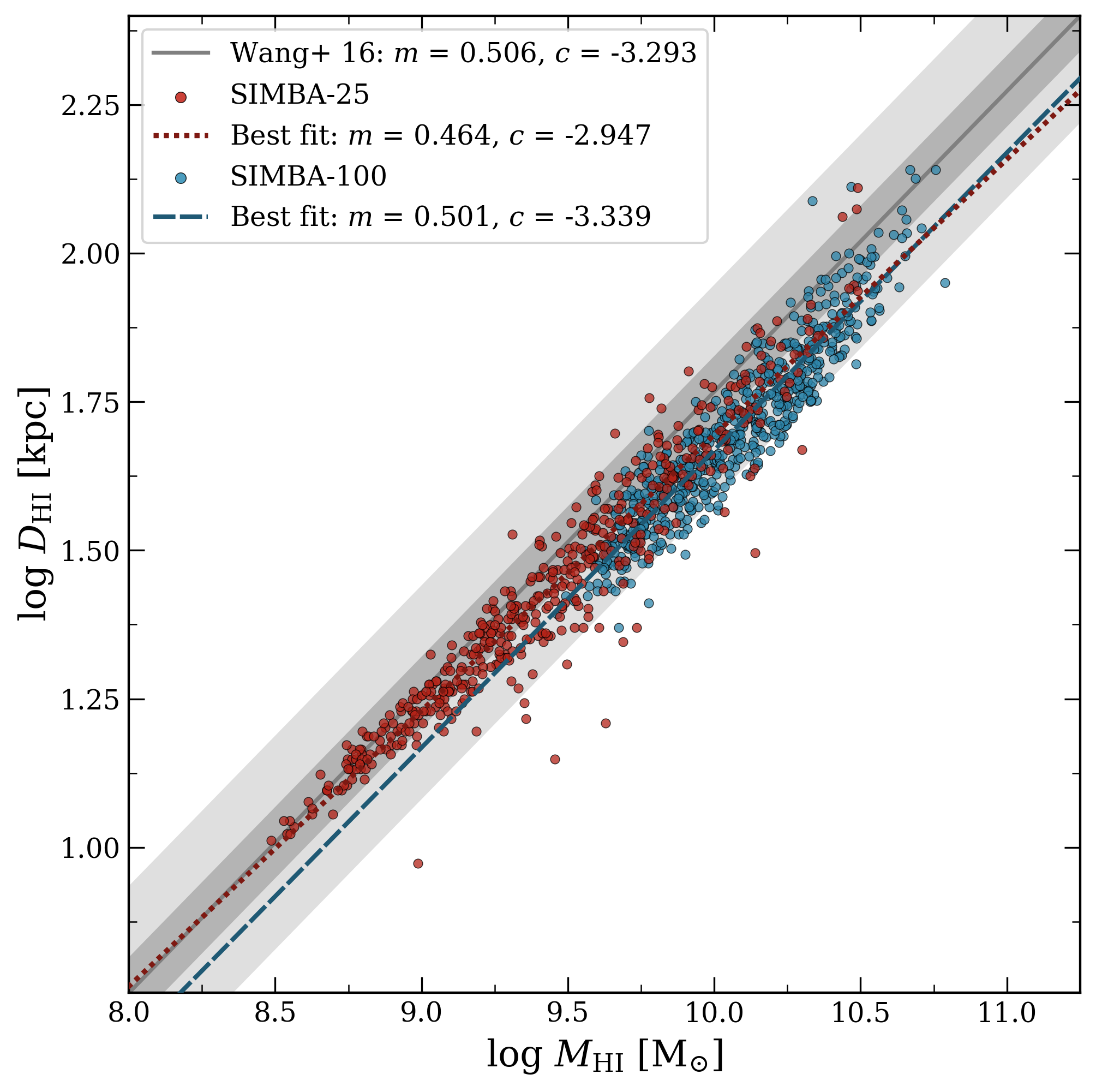}
    \caption{The \HI size-mass relation for \mbox{\simba-25} (red) and \mbox{\simba-100} (blue) galaxies. The solid grey line indicates the best fit from \protect\cite{Wang-16}, with the 1$\sigma$ and 3$\sigma$ scatter of the observational fit shown by the dark and light grey shaded regions. The red dotted and blue dashed lines represent the best fitting linear relations for \mbox{\simba-25} and \mbox{\simba-100} galaxies, respectively. For all fits shown, the slope and intercept are given in the legend.}
    \label{fig:MDrelation}
\end{figure}

We now focus on investigating the \HI size-mass relation for \mbox{\simba-25} and \mbox{\simba-100}. We correct the measured \HI diameters for beam smearing effects using

\begin{equation}
    D_{\text{\mbox{\textsc{H\,i}}}} = \sqrt{D_{\text{\mbox{\textsc{H\,i}}},0}^2-b_{\text{maj}}\times b_{\text{min}}},
\end{equation}

\noindent where $D_{\text{\mbox{\textsc{H\,i}}},0}$~($\equiv 2R_{\text{\mbox{\textsc{H\,i}}}}$) is the uncorrected \HI diameter, and $b_{\text{maj}}$ and $b_{\text{min}}$ are the major and minor axes of the beam, respectively. The corrected \HI diameters are used to construct the \HI size-mass relation, which is shown in Fig.~\ref{fig:MDrelation} along with the individual data points of each sample. The red dotted and blue dashed lines correspond to the best linear fits to the \mbox{\simba-25} and \mbox{\simba-100} data, respectively, with the fit parameters quoted in the figure legend. The solid grey line shows the best fit from \cite{Wang-16} for comparison, with the dark and light grey shaded regions representing the 1$\sigma$ and 3$\sigma$ scatter of their fit. Overall, both \simba\ samples reproduce the observed \HI size-mass relation reasonably well and the majority of galaxies lie within 3$\sigma$ of the relation. The \mbox{\simba-25} sample produces a slope that is slightly shallower than observed (0.464 compared to 0.506 in \citealt{Wang-16}, i.e. an $\sim$8~per~cent difference), resulting in galaxies that have marginally smaller \HI discs than expected at the high-\HI mass end ($M_{\text{\textsc{H\,i}}} \geq 10^{9.5}~\text{M}_{\odot}$). The \mbox{\simba-100} sample matches the observed slope remarkably well to within 1~per~cent, but the \HI discs of these galaxies are also smaller than expected given their \HI masses. It is not surprising that the \HI discs in \mbox{\simba-100} are generally more compact than observed, given the median radial profile of the \mbox{\simba-100} sample shown in Fig.~\ref{fig:medRadProf}. Nevertheless, the offset between the observed relation from \cite{Wang-16} and that obtained for \mbox{\simba-100} is relatively small ($<0.1$~dex at $M_{\text{\textsc{H\,i}}} = 10^{10}~\text{M}_{\odot}$). Despite the differences in the best linear fits obtained for \mbox{\simba-25} and \mbox{\simba-100}, we find that when only fitting the \mbox{\simba-25} galaxies with $M_{\text{\textsc{H\,i}}} \geq 10^{9.5}~\text{M}_{\odot}$, the slope and intercept measured are consistent with those of \mbox{\simba-100} within the uncertainties.

We find that the rms scatter around the relation is $0.06$~dex and $0.05$~dex for \mbox{\simba-25} and \mbox{\simba-100}, respectively, which is comparable to the observed scatter of $\sim$0.06~dex from \cite{Wang-16}. In Section~\ref{sec:DiscussHI_MD} we further discuss the \HI size-mass relations of the \simba\ samples in the context of a new study by \cite{Wang-24}, who find a slightly shallower relation when using new single dish data combined with archival \HI observations.

\section{Methodology}\label{sec:Method}

\subsection{Quantitative asymmetry measures}

A number of methods exist to quantify irregular gas distributions. Here we focus on the quantitative parameters used in this work to characterise asymmetries in the global \HI profiles and the moment-0 maps of galaxies. 

\subsubsection{Spectral asymmetries}

One of the most widely used methods of quantifying asymmetries in the global \HI profiles of galaxies was first proposed by \cite{Haynes-98}. In this method, the spectral asymmetry is calculated using the ratio of areas under the profile at velocities less than, and greater than, the systemic velocity ($v_{\text{sys}}$):

\begin{equation}
    A_{\text{1\text{D}}} = \dfrac{S_{_H}}{S_{_L}} = \dfrac{  \int^{v_{h}}_{v_{_{\text{sys}}}} \!S_{v}dv}{\int^{v_{\text{sys}}}_{v_{l}} \!S_{v}dv},
\end{equation}

\noindent where $S = \int S_{v} dv$ is the integrated flux density contained within the specified velocity bounds and $v_{l}$ ($v_{h}$) corresponds to the lower (upper) velocity measured at either 20~per~cent or 50~per~cent of the peak flux density. In the literature, the final spectral asymmetry measure, $A_{\text{1\text{D}}}$, is typically expressed as a number greater than 1, and so the inverse ratio $S_{_L}/S_{_H}$ is used instead if $S_{_H}/S_{_L}< 1$. With this definition, a value of 1 implies a perfectly symmetric global \HI profile and larger values indicate a departure from symmetry. 

We opt to use the equivalent definition used in \cite{Matthews-98}, which expresses the spectral asymmetry as a value between 0 and 1 by considering the relative flux in the two halves of the profile:

\begin{equation}
    \Aflux = \dfrac{|S_{_H}-S_{_L}|}{S_{_H}+S_{_L}} =\dfrac{ \Big| \int^{v_{h}}_{v_{_{\text{sys}}}} \!S_{v}dv - \int^{v_{_{\text{sys}}}}_{v_{l}} \!S_{v}dv \Big| }{\int^{v_{h}}_{v_{l}} \!S_{v}dv}.
\end{equation}

We use $v_{l,20}$ and $v_{h,20}$ as the velocity limits to ensure that we capture as much of the global profile as possible in the asymmetry measurement and to be consistent with previous studies \citep[e.g.][]{Espada-11,Watts-20A,Deg-20}. The profile is linearly interpolated to determine $v_{l,20}$ and $v_{h,20}$ since the limits may fall within a velocity channel. For double-horned profiles, each peak and its velocity bounds are determined separately by searching for local maxima to the left and right of the central velocity of the data cube (as an initial estimate of $v_{\text{sys}}$). If the global profile crosses the 20~per~cent flux level at multiple velocities to the left (or right) of $v_{\text{sys}}$, the innermost crossing is taken as the velocity limit. We then define $v_{\text{sys}}$ to be the midpoint of the velocity width at the 20~per~cent flux level, i.e. $v_{\text{sys}} = $~($v_{l,20}+v_{h,20}$)$/2$. The minimum number of gas elements required for the construction of the mock data cubes ensures that particle shot noise does not impact the peak flux measurements in double-horned profiles.

\subsubsection{Morphological asymmetries}\label{sec:MorphAsymm}

The subject of characterising spectral asymmetries in \HI has been investigated in greater depth than that of 2D morphological asymmetries due to the availability of data. The number of global \HI profiles, which can be obtained using single dish radio telescopes, greatly exceeds the number of spatially resolved \HI maps, which require observations with radio interferometers. As such, a number of early studies investigating \HI asymmetries in spatially resolved data adopted methods that were first introduced in optical studies. One such example is the 2D rotational asymmetry parameter. First introduced in \cite{Schade-95} and revisited in \cite{Conselice-00}, the parameter is expressed as

\begin{equation}
    A = \dfrac{\sum_{i,j} \lvert I(i,j) - I_{180}(i,j) \rvert}{2\sum_{i,j} \lvert I(i,j) \rvert},
\end{equation}

where $I(i, j)$ is the pixel flux value in the galaxy image, and $I_{180}(i,j)$ is the pixel flux value in the galaxy image rotated 180\textdegree\ about a specified rotation point. This statistic has been frequently used to study disturbed optical morphologies \citep[e.g.][]{Conselice-3,Lotz-04,Bluck-12,Abruzzo-18,Bignone-20,Thorp-21} and has more recently been extended to \HI studies \citep[e.g.][]{Holwerda-1,Holwerda-2,Giese-16,Reynolds-20A}. Possible values of $A$ range from 0 to 1, where 0 indicates a perfectly symmetric distribution.

Due to the fact that $A$ is, by definition, a flux-weighted measurement, the parameter is often dominated by asymmetric features in the bright central regions of a galaxy. Fainter features on the outskirts of galaxies, such as tidal tails, tend to have negligible contributions when calculating $A$ due to their low surface brightness. This is because the current formulation of $A$ normalises the difference between the original image and its rotated counterpart by the sum of the flux in the original image. For this reason, \cite{Lelli-14} introduced a new statistic that is more sensitive to the outer regions of the galaxy, given by:

\begin{equation}
    A_{\text{mod}} = \dfrac{1}{N} \sum\limits_{i,j}^N \dfrac{ \lvert I(i,j) - I_{180}(i,j) \rvert}{\lvert I(i,j) + I_{180}(i,j) \rvert},
\end{equation}

\noindent where $N$ is the total number of non-zero pixels in the additive image. With this formulation, the contribution of low column density asymmetries is up-weighted because the residual image is now normalised to the local flux density. The modified asymmetry parameter, hereafter $\Amod$, has been shown to reliably quantify the asymmetry in the outskirts of the \HI disc where the effects of external mechanisms are likely to be observed \citep[e.g.][]{Bilimogga-22,Deb-23,Roberts-23}. Throughout this work, we use $\Amod$ to quantify the morphological \HI asymmetries and use the centre of the moment-0 map (i.e. the minimum of the galaxy's gravitational potential) as the point around which we rotate the map. It is worth noting that our choice of rotation centre will lead to large $\Amod$ values if there is a significant offset between the galaxy's \HI disc and its potential minimum.

\cite{Bilimogga-22} investigated the impact of various observational effects, such as the spatial resolution, signal-to-noise ratio, and column density threshold implemented when measuring $\Amod$ from mock \HI observations. They found that the majority of $\Amod$ values measured using a column density threshold of $5\times 10^{19}$~cm$^{-2}$ are within $\pm0.1$ of their intrinsic $\Amod$ value irrespective of the spatial resolution. Although we are sensitive to lower column densities in our samples, we adopt the proposed threshold of $5\times 10^{19}$~cm$^{-2}$ because the $3\sigma$ column density sensitivity of the MDS is $\sim$6$\times 10^{19}$~cm$^{-2}$ over 25~km~s$^{-1}$ and we will not recover the \HI emission below this limit in the observations. The values of all pixels below the column density threshold are set to zero in the moment-0 maps when computing $\Amod$. No cuts based on signal-to-noise are necessary since our simulated moment-0 maps are noise-free. In Section~\ref{sec:Amod-Res} we examine the effect of resolution on $\Amod$ in greater detail. 

\subsection{The effect of resolution on spatially resolved asymmetries}\label{sec:Amod-Res}

Numerous studies have explored how spatial resolution impacts morphological asymmetry parameters, both in simulated \citep{Giese-16,Thorp-21,Bilimogga-22,Deg-23} and observational \citep{Reynolds-20A,Roberts-23} data, and have found that degrading the resolution of an observation leads to a decrease in the measured asymmetry due to the `smoothing' out of asymmetric features. Although high-resolution interferometric observations are desired for studying the detailed \HI morphologies of galaxies, it is also common practice to smooth observations to larger beam sizes to improve the signal-to-noise and probe the lower column density outskirts of the \HI discs. Additionally, the majority of spatially-resolved \HI detections from ongoing and future untargeted \HI surveys will be marginally resolved ($\leq 3$ beams across the major axis). It is crucial to strike a balance between determining a minimum resolution threshold that is not so strict that it significantly reduces one's sample size, while ensuring that galaxies are sufficiently well-resolved such that the measured asymmetry is still representative of the intrinsic asymmetry. As such, the minimum resolution thresholds adopted in the studies mentioned above are somewhat subjective and vary across the studies.

We approach the subject of resolution effects from a slightly different angle, by examining whether the differences in spatial resolution across a sample can result in a biased relationship between \HI mass and $\Amod$. \cite{Glowacki-22} observed a weak trend of decreasing spectral asymmetry with increasing \HI mass, but this has not been investigated with morphological asymmetry measurements. To quantify how well resolved a galaxy is, we sum up the total number of pixels above $5\times 10^{19}$~cm$^{-2}$ in the moment-0 map, and divide by the number of pixels in the $15\arcsec$ circular beam to get the number of beams in the area of the galaxy, $N_{\text{beams}}$. We use this approach instead of calculating the number of beams across the major axis to account for the fact that a face-on galaxy has more resolution elements in its area than an edge-on galaxy with the same measured major axis.

\begin{figure*}
 \centering\includegraphics[width=\linewidth]{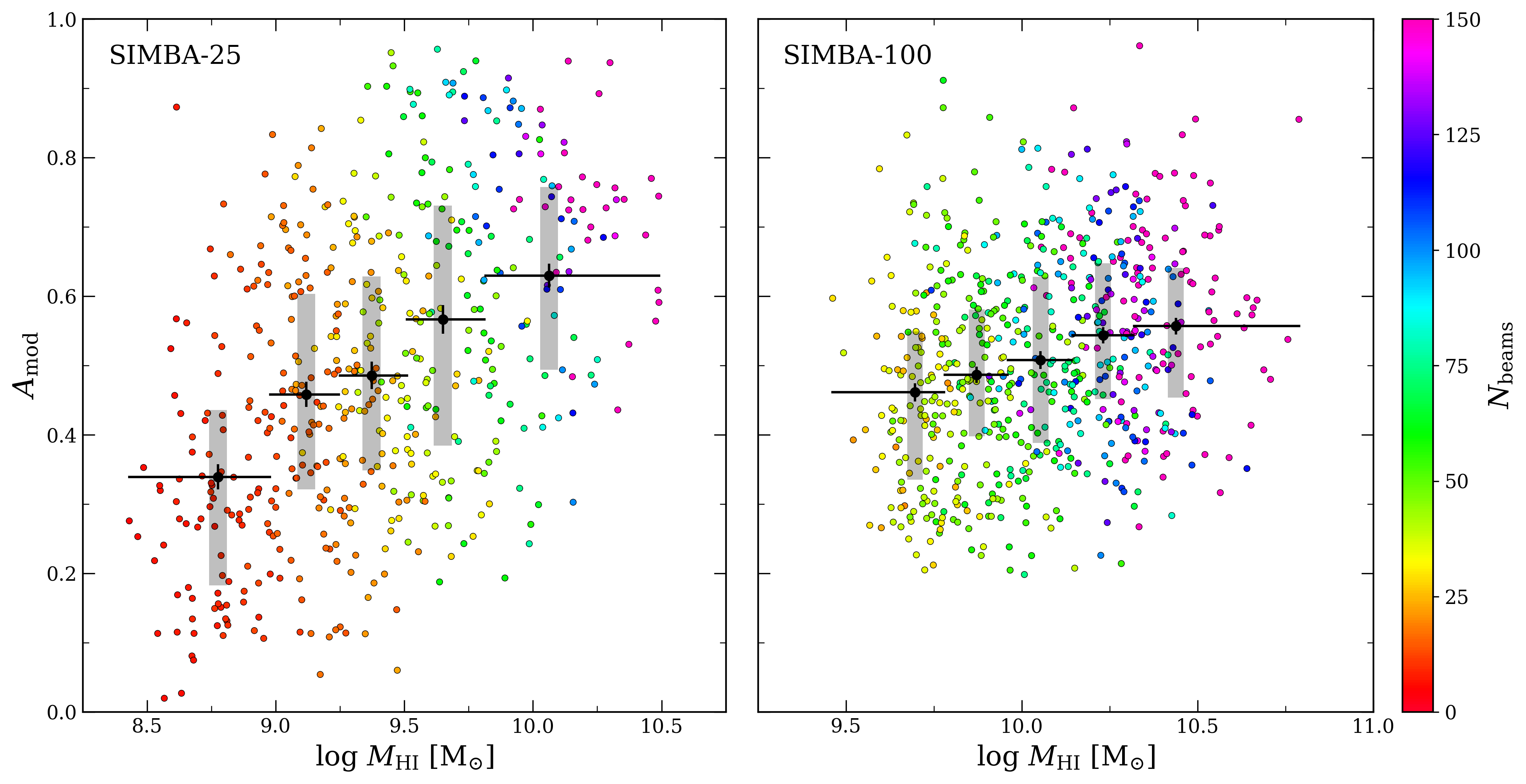}
 \caption{Scatter plot of morphological asymmetry ($\Amod$) as a function of \HI mass for \mbox{\simba-25} (left) and \mbox{\simba-100} (right) galaxies. Here, $\Amod$ is calculated using the $15\arcsec \times 15\arcsec$ resolution moment-0 maps with an applied column density threshold of $5\times 10^{19}$~cm$^{-2}$. The colour of individual data points shows the number of beams in the area of the galaxy. The large black circle markers indicate the mean $\Amod$ value in bins of increasing \HI mass, with the bin widths shown by the horizontal error bars. The vertical error bars show the standard error on the mean ($1\sigma/\sqrt{N}$), where $\sigma$ is the standard deviation and $N$ is the number of galaxies in the bin. The shaded regions show the interquartile ($25^{\text{th}}-75^{\text{th}}$~percentile) range.}
 \label{fig:MHI-AmodUncorrected}
\end{figure*}

In Fig.~\ref{fig:MHI-AmodUncorrected} we show $\Amod$ as a function of \HI mass for the \mbox{\simba-25} (left panel) and \mbox{\simba-100} (right panel) galaxies. The data points are colour-coded by $N_{\text{beams}}$ calculated at $15\arcsec$ resolution. Each sample is divided into 5 bins of increasing \HI mass, with varying bin widths to ensure that each bin in the respective sample has approximately the same number of galaxies. The mean $\Amod$ value of each bin is illustrated by the black circle markers, with the vertical error bars and shaded regions representing the standard error on the mean ($1\sigma/\sqrt{N}$) and the interquartile range, respectively. The horizontal error bars represent the bin widths. 

\begin{figure*}
 \centering\includegraphics[width=\linewidth]{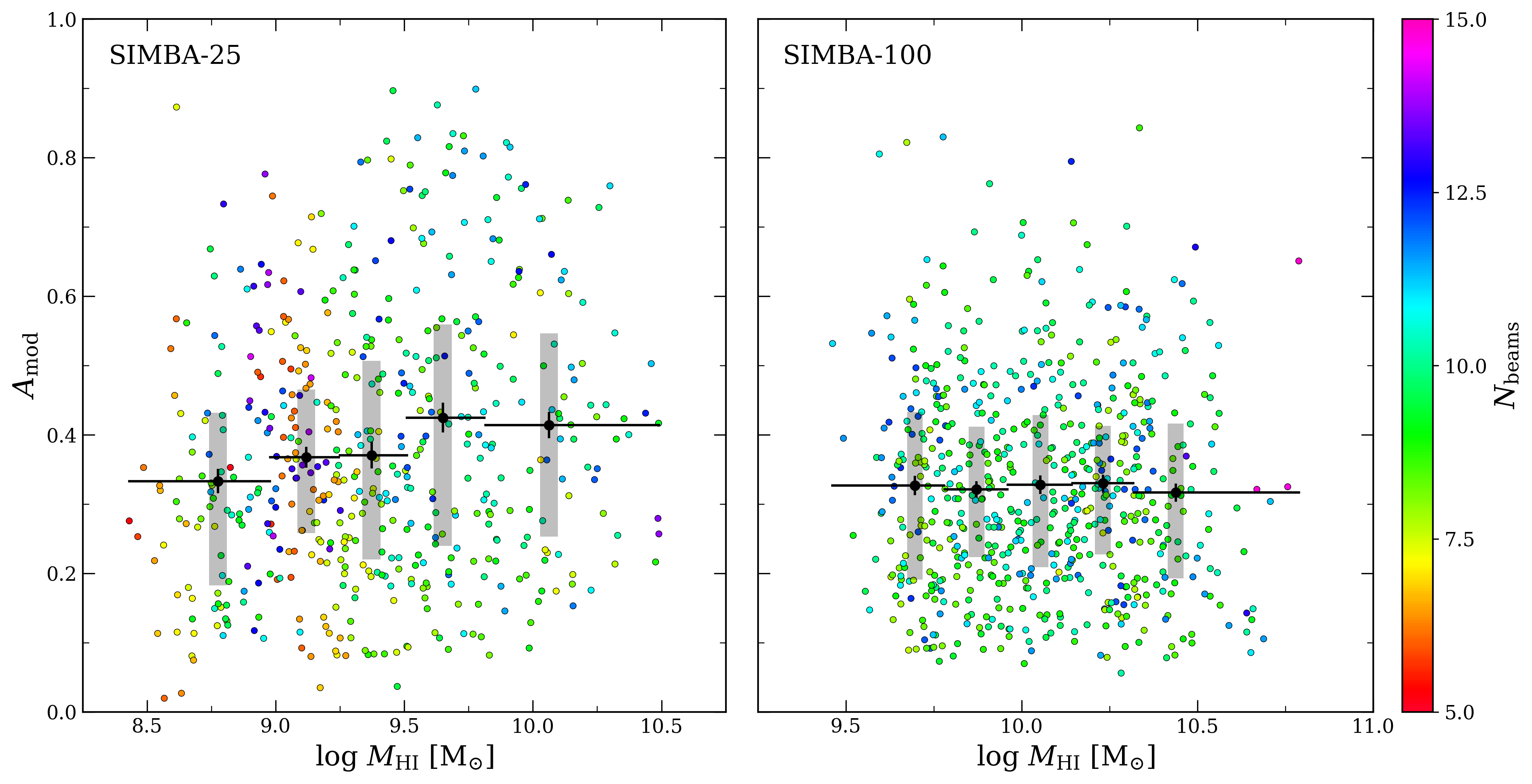}
 \caption{Same as in Fig.~\ref{fig:MHI-AmodUncorrected}, except $\Amod$ is calculated using the resolution at which a galaxy is resolved by approximately 10 beams in its area. The colour of individual data points shows the number of beams in the area of the galaxy at the resolution used when calculating $\Amod$.}
 \label{fig:MHI-AmodCorrected}
\end{figure*}

Despite considerable scatter within each bin, shown by the spread in the individual data points as well as the interquartile range, it is evident that $\Amod$, when measured at a fixed resolution, tends to increase with increasing \HI mass. A Spearman rank test of \HI mass and $\Amod$ gives a correlation coefficient of $r_{s} = 0.46$ ($p$-value $< 0.001$) for \mbox{\simba-25} and $r_{s} = 0.25$ ($p$-value $< 0.001$) for \mbox{\simba-100}, suggesting a weak but statistically significant correlation. This trend is less pronounced in the \mbox{\simba-100} sample, which could be partly due to the compactness of the \HI discs in the sample (see Section~\ref{sec:DataVerification}) as well as the fact that the \mbox{\simba-100} sample spans a smaller \HI mass range than the \mbox{\simba-25} sample. 

From the colour gradient observed in both panels of Fig.~\ref{fig:MHI-AmodUncorrected}, it is apparent that more massive \HI discs are more resolved. This is not surprising since we have shown that both \simba\ samples adhere to the \HI size-mass relation. However, this raises the question of whether the positive relation between \HI mass and $\Amod$ is a consequence of resolution. The median $N_{\text{beams}}$ of galaxies in the lowest and highest mass bins of the \mbox{\simba-25} sample are $\sim$9.5 and $\sim$104.4, respectively. In the case of \mbox{\simba-100}, the median $N_{\text{beams}}$ of galaxies in the lowest and highest mass bins are $\sim$37.4 and $\sim$159.1, respectively. This means that the highest \HI mass galaxies can be 4-11 times more resolved than the lowest \HI mass galaxies. Furthermore, we do not find any galaxies in the \mbox{\simba-100} sample with $\Amod \leq 0.2$, and those galaxies with low $\Amod$ values in the \mbox{\simba-25} sample tend to be resolved by less than $\sim$25 beams. 

To assess the dependence of this observed trend on resolution, we reduce the number of resolution elements across the moment-0 maps by smoothing to larger circular beam sizes of 30, 45, 60, 90, and 120\arcsec. We calculate $N_{\text{beams}}$ at each resolution in the same manner as before, by summing the pixels in the moment-0 map above $5\times 10^{19}$~cm$^{-2}$ and dividing by the pixel area of the beam. Throughout this exercise, the pixel size is fixed at $5\arcsec \times5 \arcsec$\footnote{We tested whether downsampling the pixels (by increasing their size) in the smoothed data affects the Amod values and found no systematic difference, despite slightly increased scatter at low Amod values.}. Although we become more sensitive to lower column densities when degrading the resolution, we maintain the same column density threshold to ensure a consistent comparison. We identify the resolution at which a galaxy is resolved by approximately 10 beams and recalculate $\Amod$ at this resolution. This is done instead of enforcing that a galaxy be resolved by at least 10 beams, as the latter results in a broader distribution of $N_{\text{beams}}$ for both samples. With this approach, all galaxies are resolved by at least 5 beams, but no more than 15 beams. In Appendix~\ref{sec:AppendixMom0} we show the moment-0 maps of a few galaxies, at the initial $15\arcsec$ resolution and smoothed to the resolution at which $N_{\text{beams}}\approx10$, along with their global \HI profiles. 

Fig.~\ref{fig:MHI-AmodCorrected} shows $\Amod$ as a function of \HI mass for both samples when $N_{\text{beams}}$ has been standardised. The colour gradient observed in Fig.~\ref{fig:MHI-AmodUncorrected} is largely absent in Fig.~\ref{fig:MHI-AmodCorrected} as galaxies now have similar $N_{\text{beams}}$ by construction. There is a small vertical band of galaxies with $N_{\text{beams}} \geq 12$ around $M_{\text{\textsc{H\,i}}}\approx 10^{9}~\text{M}_{\odot}$ in the \mbox{\simba-25} sample, but these galaxies have a range of $\Amod$ values and are not preferentially more asymmetric despite being marginally more resolved. Degrading the resolution of the moment-0 maps results in a clear shift towards lower $\Amod$ values at higher \HI masses in both samples. The low $\Amod$ `floor' of the \mbox{\simba-100} sample that is present at $\Amod \approx 0.2$ in the right panel of Fig.~\ref{fig:MHI-AmodUncorrected} is not observed in Fig.~\ref{fig:MHI-AmodCorrected} and a larger fraction of galaxies from both samples populate the low $\Amod$ region of the figure. Reassuringly, we still observe highly asymmetric galaxies ($\Amod \geq 0.6$) in both samples, which gives us confidence that the galaxies are sufficiently resolved and that asymmetric features have not been completely washed out by degrading the resolution. 

There is no discernible trend between $\Amod$ and \HI mass in the \mbox{\simba-100} sample when galaxies are resolved by approximately the same number of beams. We find that the trend persists in the \mbox{\simba-25} sample according to the Spearman rank test ($r_{s} = 0.15$, $p$-value $< 0.001$) but it is considerably weaker than previously observed, indicating that the previous trend between $\Amod$ and \HI mass was exaggerated due to the effect of resolution. If we only consider the \HI mass range $10^{9.5}~\text{M}_{\odot} \leq M_{\text{\mbox{\textsc{H\,i}}}} \leq 10^{10.5}~\text{M}_{\odot}$, where the two samples overlap, we find no correlation between $\Amod$ and \HI mass in the \mbox{\simba-25} sample ($r_{s} = 0.03$, $p$-value $= 0.67$). This suggests that the conflicting results between the two samples is due to the limited \HI mass range of the \mbox{\simba-100} sample.

Overall, these findings demonstrate that resolution effects can still persist if only a minimum resolution threshold is applied, and that failing to account for these effects can produce misleading results. If a sample spans a large \HI mass range, one must either bin galaxies into different groups based on $N_{\text{beams}}$ or \HI mass to compare $\Amod$ values in a fair and consistent manner, or degrade the resolution of galaxies as has been done in this work. We opt for this approach because it allows us to utilise the full sample when investigating correlations between $\Amod$ and other galaxy properties. In the remainder of this work, we use the $\Amod$ values measured at the resolution at which $N_{\text{beams}}\approx10$.

\section{Physical drivers of H I asymmetry}\label{sec:DriversofAsymm}

As previously mentioned, disturbances in the \HI reservoirs of galaxies can be caused by various physical processes that lead to the depletion or replenishment of a galaxy's \HI content. A number of studies utilising cosmological simulations have focused on the dependence of environment on spectral asymmetries \citep[e.g.][]{Watts-20B,Manuwal-22} and have found that while satellite galaxies, as a population, are typically more asymmetric than central galaxies, environmental processes are not the sole driver of the observed asymmetries in the global \HI profiles. 

We take advantage of the ability to trace a galaxy's recent dynamical history in the simulations to investigate galaxy-galaxy interactions and mergers as the possible origins of \HI asymmetries in our samples. We consider a total of 11 snapshots between the redshift range $0.0167 \leq z \leq 0.192$, corresponding to a period of $\sim$2.2 Gyr. Previous estimates of the observability timescales of tidal features and disturbed morphologies range from $\sim$0.2$-$2~Gyr, where the timescale depends on the mass ratio, gas fractions and relative orientations of the interacting galaxies \citep[e.g.][]{Lotz-10A,Lotz-10B,Holwerda-11-3,Whitney-21}. Thus, we do not expect that any signatures of an interaction or merger that took place beyond $z\sim0.2$ will be sustained until $z=0.0167$, because a galaxy will likely have had enough time to regularize its \HI disc over multiple revolutions. 

We use \caesar\ to track the galaxies in our two samples across the snapshots of interest. For each target galaxy in the $z=0.0167$ snapshot, its most massive progenitor in the previous snapshots is identified as the galaxy that contains the most shared star particle IDs. For each snapshot, we then obtain the stellar mass, \HI mass and baryonic mass ($M_{\text{baryon}} = M_{\star} +$~1.33$M_{\text{\textsc{H\,i}}}$) of the target galaxy to determine its mass growth history, as well as the 3D separations ($r_{\text{sep}}$) between the target galaxy and its neighbouring galaxies. 

To identify galaxies that have undergone a merger within the specified time period, we compute the fractional variation in a galaxy's stellar and baryonic mass using

\begin{equation}
    \mathcal{R} = \dfrac{M_{t}-M_{t-1}}{M_{t-1}},
\end{equation}

where $M_{t}$ is the galaxy's mass at a given snapshot and $M_{t-1}$ is its mass at the previous snapshot. We define a merger as a fractional increase in the stellar and baryonic mass ($\mathcal{R}_{\star}$ and $\mathcal{R}_{\text{baryon}}$, respectively) of $\geq 5$~per~cent. Our reason for considering $\mathcal{R}_{\star}$ in addition to $\mathcal{R}_{\text{baryon}}$ is to reduce misidentifications at the low mass end, where \HI\!$-$rich dwarf galaxies may experience large jumps in their baryonic mass due to gas inflows and be incorrectly identified as mergers. Following a similar approach as \cite{Montero-19}, we estimate the fractional stellar mass increase that can be attributed to star formation ($\mathcal{R}_{\text{SF}}$) using the SFRs provided in the \caesar\ catalogues and the time difference between the two snapshots where the mass increase occurs, and require that $\mathcal{R}_{\star}-\mathcal{R}_{\text{SF}} \geq 5$~per~cent to ensure that the observed stellar mass increase cannot be explained solely by enhanced star formation. Lastly, we require that prior to the target galaxy's mass increase, the closest neighbouring galaxy must have $r_{\text{sep}}\leq50$~kpc, and that the baryonic mass ratio between the neighbouring galaxy and the target galaxy ($M_{\text{baryon}}$ ratio) be $\geq 5$~per~cent. 

For recent close encounters, we also consider interactions between galaxies with $M_{\text{baryon}}$ ratio $\geq 5$~per~cent, but adopt a minimum $r_{\text{sep}}$ of 100~kpc. Although interaction-driven effects such as enhanced star formation have been observed in close galaxy pairs with projected separations of $\sim$150~kpc, a stellar mass ratio $\geq 10$~per~cent is typically used in the existing literature \citep[e.g.][]{Patton-13,Moreno-21}. We opt for 100~kpc as a compromise for the lower $M_{\text{baryon}}$ ratio adopted in this work, which is chosen to be consistent with how mergers are identified. 

With these criteria, we identify 136 galaxies (78 and 58 from \mbox{\simba-25} and \mbox{\simba-100}, respectively) that have undergone a merger, 189 galaxies (106 and 83) that have had a close interaction, and 832 galaxies (329 and 503) that have remained relatively isolated over the preceding $\sim$2.2~Gyr. For simplicity, we hereafter refer to these categories as merged, interacted, and isolated, respectively. Based on our criteria, it is possible that galaxies in the isolated category may have experienced interactions with neighbouring galaxies that meet the minimum $r_{\text{sep}}$ condition but not the $M_{\text{baryon}}$ ratio condition, or vice versa. To examine if these selection criteria impact our results, we define a refined isolated category consisting of galaxies that have had no neighbouring galaxy (regardless of mass) within 100~kpc and no significant neighbouring galaxy (with $M_{\text{baryon}}$ ratio $\geq 5$~per~cent) within 500~kpc over the considered time period. The unmasked moment-0 maps of the refined isolated galaxies were also visually inspected to ensure the absence of extraneous \HI emission. Of the 832 isolated galaxies, 306 galaxies (94 and 212) belong to the refined isolated category. 

For the remainder of this work, we have combined the samples from \mbox{\simba-25} and \mbox{\simba-100} to increase the number of galaxies per category for better statistics, and to enable us to investigate trends between \HI asymmetry and baryonic mass. In the following three sections, we study the distributions of the asymmetry parameters for the three categories, how \HI spectral and morphological asymmetries compare, and whether \HI asymmetry is correlated with baryonic mass.

\subsection{The relationship between \texorpdfstring{\HI}~asymmetry and recent dynamical history}\label{sec:AsymmDist}

\begin{figure}
	\includegraphics[width=\columnwidth]{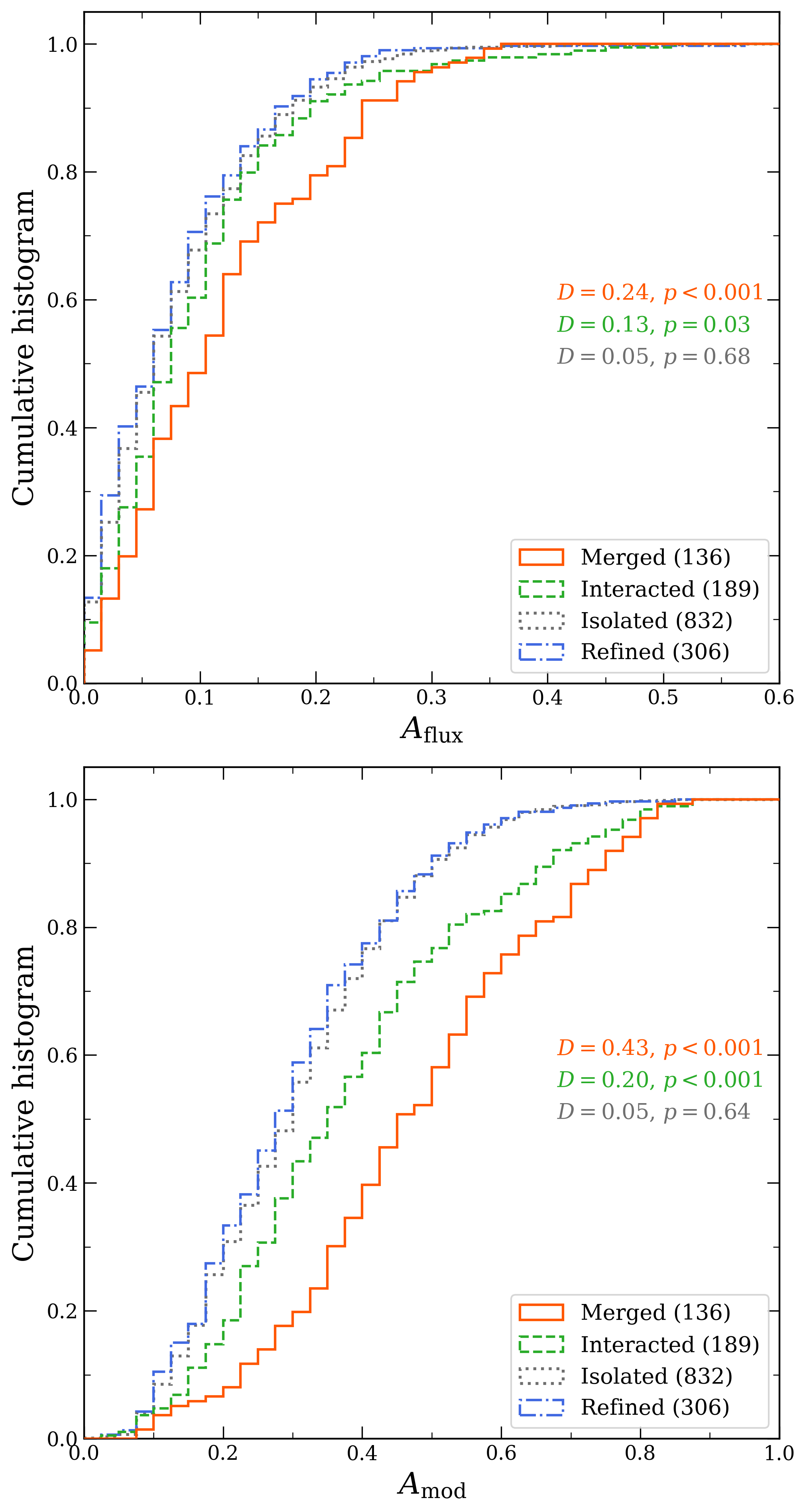}
    \caption{Comparison between the cumulative histograms of the spectral asymmetry ($\Aflux$, top) and morphological asymmetry ($\Amod$, bottom) for galaxies that have undergone a merger (mass ratios $>1$:20) within the last $\sim$2.2~Gyr (orange, solid), galaxies that have experienced an interaction (green, dashed) and galaxies that have remained relatively isolated (grey, dotted) within the same time period. In both panels, the refined isolated category's cumulative histogram is shown by the blue, dash-dotted histogram. The results of the two-sample Kolmogorov-Smirnov tests comparing the isolated, interacted and merged categories to the refined isolated category are given in both panels.}
    \label{fig:CHist_AmodAfl}
\end{figure}

\begin{table}
\centering
\caption{The median (P$_{50}$) and interquartile range (P$_{75} - $~P$_{25}$) of $\Amod$ and $\Aflux$ measured for the different categories.}
\label{table:MeanStd3Cats}
\begin{tabular}{lcccc}
\hline
 & \multicolumn{2}{c}{Aflux} & \multicolumn{2}{c}{Amod} \\
 & P$_{50}$ & P$_{75} - $~P$_{25}$ & P$_{50}$ & P$_{75} - $~P$_{25}$ \\ \hline
Refined & 0.065 & 0.092 & 0.289 & 0.212 \\
Isolated & 0.067 & 0.097 & 0.307 & 0.217 \\
Interacted & 0.082 & 0.088 & 0.364 & 0.266 \\
Merged & 0.106 & 0.126 & 0.471 & 0.265 \\ \hline
\end{tabular}
\end{table}

In Fig.~\ref{fig:CHist_AmodAfl} we show the cumulative distributions of $\Aflux$ (top panel) and $\Amod$ (bottom) for galaxies in the isolated, refined isolated, interacted, and merged categories (grey dotted,blue dash-dotted, green dashed, and orange solid histograms, respectively). For ease of comparison, we provide the median and interquartile range of $\Aflux$ and $\Amod$ for each category in Table~\ref{table:MeanStd3Cats}. Before comparing the cumulative distributions of different categories, it is important to note the differences in the range of values of $\Aflux$ and $\Amod$. Although both asymmetry parameters can theoretically have a maximum value of 1, we do not measure an $\Aflux$ value $> 0.6$ in our sample, while $\Amod$ covers a slightly larger range of values from $0-0.9$. It is also apparent that the distributions of $\Aflux$ and $\Amod$ are different. The cumulative distributions of $\Aflux$ for all categories are positively skewed, indicating that a relatively large fraction of galaxies have low spectral asymmetry and a small fraction of galaxies have high spectral asymmetry, whereas the $\Amod$ distributions are generally broader and do not exhibit the same behaviour. The interquartile range of $\Aflux$ for all categories is also smaller than that of $\Amod$ by a factor of $\sim$2$-3$, confirming the larger spread in $\Amod$ compared to $\Aflux$.

Starting with the $\Aflux$ distributions of the different categories, we find that merged galaxies have the highest mean asymmetry, followed by interacted galaxies, with refined isolated galaxies having the lowest mean asymmetry and smallest standard deviation (Table~\ref{table:MeanStd3Cats}). This trend is emphasised in the figure, where the cumulative histogram of the merged category has a shallower slope and a lower cumulative fraction than that of the refined isolated category at a given $\Aflux$ value, confirming that the merged category is, on average, more asymmetric. The differences between the cumulative histograms of the refined isolated and interacted categories are subtle and only become apparent at $\Aflux > 0.2$, where the interacted category has an extended tail towards high $\Aflux$ values. We observe a changeover at $\Aflux \approx 0.32$, where the interacted category has a lower cumulative fraction than the merged category at high $\Aflux$ values. However, this changeover occurs at a cumulative fraction of $\sim$0.95, indicating that the interacted category is only more asymmetric than the merged category for a small fraction ($< 5$~per~cent) of the most asymmetric global \HI profiles. We find that the isolated category is only slightly more asymmetric than the refined isolated category, but has a comparable standard deviation and slightly smaller standard error on the mean due to the larger sample size.

As mentioned previously, it has become common practice to characterise the prevalence of spectral asymmetries in a sample by determining the fraction of galaxies with measured asymmetry values above a certain threshold. Recent studies utilising the $A_{\text{1\text{D}}}$ index from \cite{Haynes-98} adopted threshold values of 1.26 and 1.39, corresponding to deviations greater than $2\sigma$ and $3\sigma$ from the mean of the asymmetry distribution of isolated AMIGA galaxies as determined by \cite{Espada-11}. It is possible to get the equivalent thresholds expressed as a number between 0 and 1 using the conversion $\Aflux = (A_{\text{1\text{D}}}-1)/(A_{\text{1\text{D}}}+1)$, which gives us limits of $\Aflux > 0.12$ and $> 0.16$. We use these values to calculate the relative fractions of asymmetric global \HI profiles in the different categories and estimate the uncertainties in the fractions using Poisson error propagation. The results are quoted in Table~\ref{table:AfluxRates}. The asymmetric fractions of galaxies with $\Aflux > 0.12$ ($> 0.16$) in the refined isolated, isolated, interacted and merged categories are 26~per~cent (14~per~cent), 28~per~cent (15~per~cent), 33~per~cent (17~per~cent), and 46~per~cent (29~per~cent), respectively. Isolated galaxies have comparable, but slightly higher, asymmetric fractions to the refined isolated galaxies. In contrast, the merged category has significantly higher fractions than both isolated categories, and the interacted category has asymmetric fractions that are consistent with both isolated categories, within the uncertainties. We compare these results with the existing literature in Section~\ref{sec:Discuss_AfluxRate}. 

\begin{table}
\centering
\caption{Comparison of the relative fractions of galaxies with $\Aflux$ values above a certain threshold in each category. }
\label{table:AfluxRates}
\begin{tabular}{lccc}
\hline
     &   $N$   & $\Aflux > 0.12$ & $\Aflux > 0.16$ \\
     &     & (per~cent) & (per~cent) \\ \hline
Refined   &  306  & $26 \pm 3$             & $14 \pm 2$             \\
Isolated   &  832  & $28 \pm 2$             & $15 \pm 1$             \\
Interacted   &  189  & $33 \pm 4$            & $17 \pm 3$            \\
Merged   &  136  & $46 \pm 7$             & $29 \pm 5$             \\ \hline
\end{tabular}
\end{table}

The differences between the categories become even more pronounced when considering the cumulative distributions of $\Amod$. At low $\Amod$ values, the interacted category's cumulative histogram is narrowly offset from both isolated categories and has a similar slope, before diverging at $\Amod\approx0.3$. In comparison, the merged category's cumulative histogram is more clearly separated and systematically offset towards higher asymmetry than both the isolated and interacted categories. Again, the cumulative histograms of the isolated and refined isolated categories are marginally offset from each other. 

To quantify the differences between the categories' distributions of $\Aflux$ and $\Amod$, we perform two-sample Kolmogorov-Smirnov (KS) tests comparing the isolated, interacted and merged categories to the refined isolated category, with the results quoted in the panels of Fig.~\ref{fig:CHist_AmodAfl}. Despite the small offsets observed between the cumulative distributions of the isolated and refined isolated categories, given the $p$-values we cannot reject the null hypothesis that they have the same distribution. We also perform KS tests between the refined isolated category and the residual subset of the isolated category, and still obtain $p$-values $>0.1$. These results suggest that including galaxies that have had close encounters with companion galaxies below our specified baryonic mass threshold, or encounters with more massive galaxies beyond $r_{\text{sep}} \geq 100$~kpc, has a negligible impact on the $\Aflux$ and $\Amod$ distributions of galaxies in the isolated category. To avoid redundancy, the remaining results do not include the refined isolated category since the category's asymmetry distributions are statistically indistinguishable from that of the isolated category, and the latter has a larger sample size.

We find that the distributions of $\Amod$ and $\Aflux$ of merged galaxies are statistically different from those of the isolated galaxies. For both the interacted and merged categories, the KS test statistic $D$ returned for $\Amod$ is greater than that for $\Aflux$, confirming that there is a larger distinction between the $\Amod$ distributions of different categories than there is for the $\Aflux$ distributions. Furthermore, while the $\Amod$ distribution of interacted galaxies is statistically different to that of the isolated category, this is not the case for $\Aflux$. Therefore, it seems that the \HI morphologies of the interacted galaxies are typically more asymmetric than the \HI morphologies of the isolated galaxies, but their global \HI profiles do not show the same clear differences. Similarly, we also find a statistically significant difference between the $\Amod$ distributions of the interacted and merged categories ($p$-value $< 0.001$), but no strong difference between their $\Aflux$ distributions ($p$-value $= 0.03$). 

\begin{figure*}
 \centering\includegraphics[width=\linewidth]{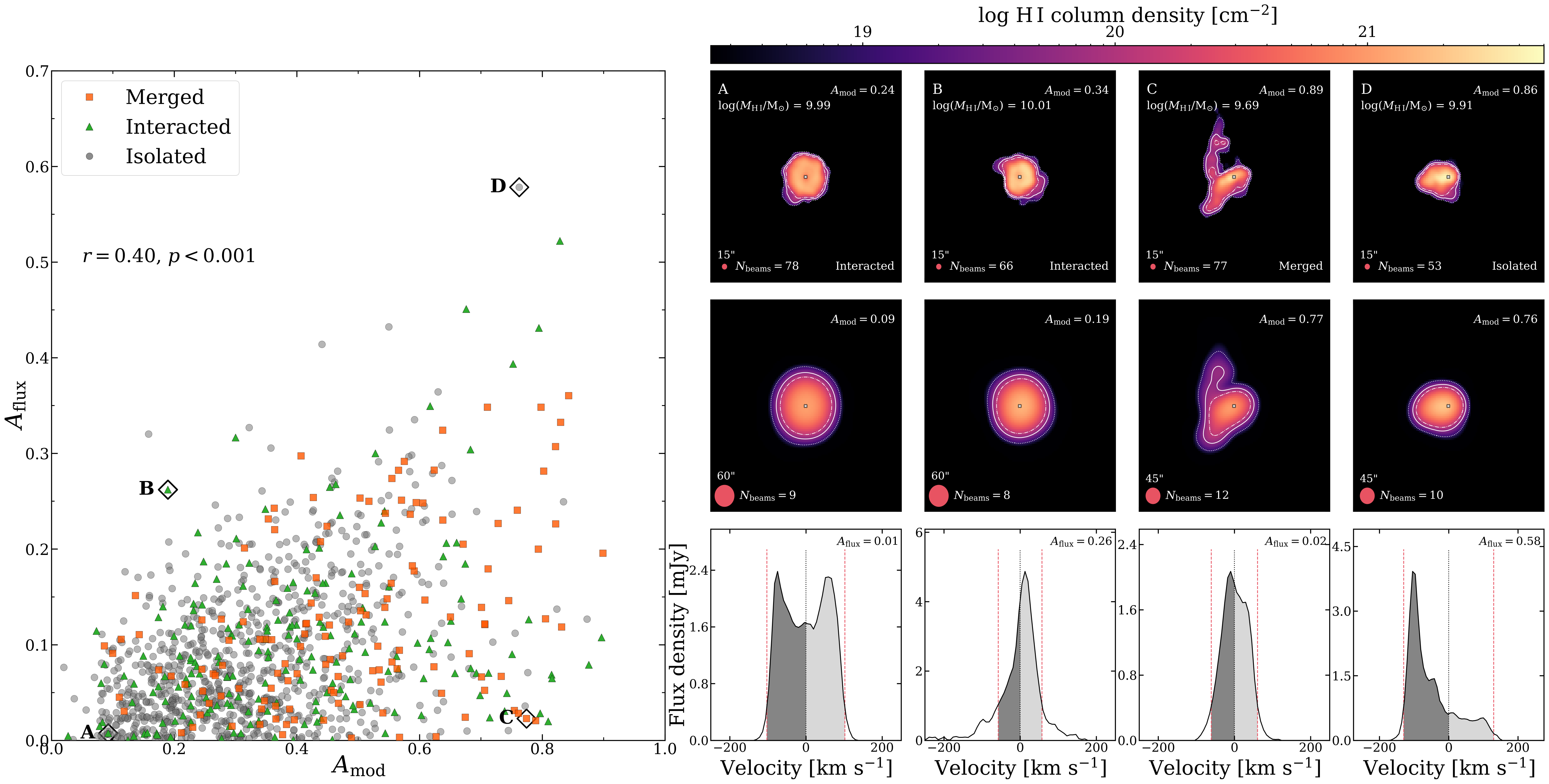}
 \caption{Left: Spectral asymmetry ($\Aflux$) plotted against morphological asymmetry ($\Amod$) for the isolated (grey circles), interacting (green triangles), and merged (orange squares) galaxies in our sample. The Spearman correlation coefficient and associated $p$-value using the full sample are quoted in the panel. Right: The moment-0 maps (top and middle rows) and global \HI profiles (bottom) of four galaxies located in different regions of the left panel. The moment-0 maps are shown at $15\arcsec$ resolution (top row) and the resolution at which $N_{\text{beams}}\approx10$ (middle). Each panel showing the moment-0 maps is 200~kpc across and centred on the potential minimum of the galaxy, which is shown by a white cross. The white dash-dotted, solid, and dotted contours are at column densities of $1,5, \text{ and }12.5\times 10^{19}$~cm$^{-2}$. In the bottom row, the dashed pink lines represent $v_{l,20}$ and $v_{h,20}$, and the black dotted line represents $v_{\text{sys}}$. Galaxies A (ID: 40) and C (ID: 1997) are from \mbox{\simba-25} and galaxies B (ID: 3435) and D (ID: 34668) are from \mbox{\simba-100}. The logarithm of the \HI mass and the category of each galaxy are given in the top panels. The measured $\Amod$ and $\Aflux$ values are quoted in the top right corner of the relevant panel.}
 \label{fig:AmodAfluxPlot}
\end{figure*}

\subsection{Comparison between asymmetry measures}\label{sec:AmodAflCor}

We turn to a direct comparison of $\Aflux$ and $\Amod$ on a galaxy-by-galaxy basis to investigate the differences in the distributions of the two parameters in more detail and explore the correlation between the parameters. Because the majority of \HI detections from upcoming \HI surveys will be spatially unresolved but spectrally resolved, understanding whether a correlation exists between spectral and morphological asymmetries will enable us to determine if a galaxy's global \HI profile can be used to infer the state of its underlying \HI disc. Recent studies by \cite{Reynolds-20A} and \cite{Bilimogga-22} have examined the correlations between various spectral and morphological \HI asymmetries and produced conflicting results. The tension between these studies' findings could be due to numerous factors, such as the parameters used to quantify the asymmetries, the differences in the mass ranges of the samples and the relatively small ($N < 200$) sample sizes, as well as the treatment of observational effects. 

The left panel of Fig.~\ref{fig:AmodAfluxPlot} shows $\Aflux$ against $\Amod$ for isolated, interacted and merged galaxies (grey circle, green triangle and orange square markers, respectively). For comparison, the right half of Fig.~\ref{fig:AmodAfluxPlot} shows the moment-0 maps and global \HI profiles of four galaxies located in different regions of the scatter plot (see the figure caption for more details). While the merged and interacted categories were previously found to have higher mean asymmetries than the isolated category, this does not translate to any category occupying a specific region of the figure as there is considerable overlap between the three categories at various values of $\Aflux$ and $\Amod$. Remarkably, there are still isolated galaxies with high $\Aflux$ and/or high $\Amod$ values. We discuss possible explanations for this in Section~\ref{sec:Discuss_AfluxRate}.

It is not surprising that the majority of galaxies from all three categories fall below $\Aflux = 0.16$, given the results in Table~\ref{table:AfluxRates}. However, it is apparent that these galaxies with low $\Aflux$ values can still have a large range of $\Amod$ values. On the other hand, the majority of galaxies with high $\Aflux$ values $>0.3$ tend to have high $\Amod$ values $>0.6$. This suggests that galaxies with highly asymmetric global \HI profiles are likely to have disturbed \HI discs, but a symmetric global profile is not necessarily indicative of a symmetric \HI disc. Because the shape of a galaxy's global \HI profile is dependent on both the kinematics of the galaxy and the total mass distribution of \HI\!, it is possible that distortions in both properties may conspire to produce a symmetric global \HI profile. Furthermore, the shape of the global \HI profile is also affected by viewing angle and inclination, which will in turn affect the measured asymmetry \citep[e.g.][]{Deg-20}. It is also apparent that the top left region of the left panel of Fig.~\ref{fig:AmodAfluxPlot}, where galaxies with low $\Amod$ and high $\Aflux$ values would reside, is virtually empty compared to the other regions of the figure. In general, galaxies with low $\Amod$ values tend to have relatively symmetric global \HI profiles, indicating that these galaxies must also have fairly regular kinematics. A Spearman rank test performed using the full sample gives a correlation coefficient of $r_{s} = 0.40$ ($p$-value $< 0.001$), indicating a weak, but statistically significant, correlation between $\Aflux$ and $\Amod$. We compare these results with previously published studies in Section~\ref{sec:Discuss_AfluxRate}. 

\subsection{The relationship between \texorpdfstring{\HI}~asymmetry and baryonic mass}\label{sec:Mbar-HIasymm}

\begin{figure*}
 \centering\includegraphics[width=\linewidth]{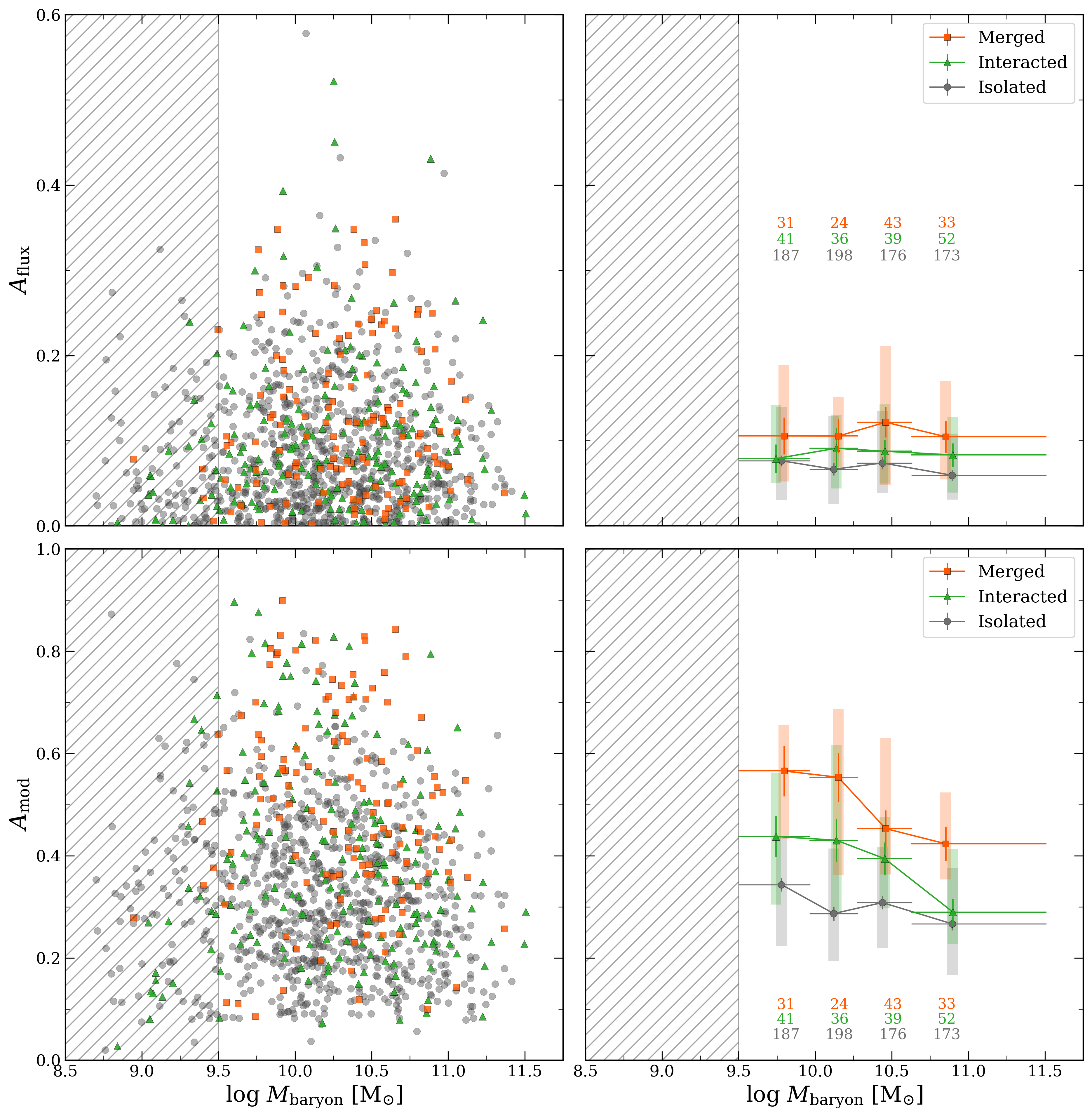}
 \caption{\HI spectral asymmetry ($\Aflux$, top) and morphological asymmetry ($\Amod$, bottom) as a function of baryonic mass. The left panels show the individual data points from the isolated (grey circles), interacted (green triangles) and merged (orange squares) categories. In the right panels, the markers indicate the median \HI asymmetry of each category in bins of increasing baryonic mass, with the bin widths shown by the horizontal error bars. The vertical error bars and shaded regions show the standard error on the median and the interquartile ($25^{\text{th}}-75^{\text{th}}$~percentile) range, respectively. Galaxies that lie in the grey hatched region (with $M_{\text{baryon}} < 10^{9.5}~M_{\odot}$) are excluded from the binned median trends. The number of galaxies in each category is quoted per bin.}
 \label{fig:Asymm-MbarFS}
\end{figure*}

Having analysed the cumulative distributions of $\Amod$ and $\Aflux$ for our sample, along with the correlation between the two asymmetry parameters, we now explore possible trends between \HI asymmetry and baryonic mass. Previous studies have investigated correlations between \HI spectral asymmetry and stellar mass (as a proxy for halo mass), albeit with inconclusive results on whether a mass dependence exists \citep[e.g.][]{Watts-20A,Reynolds-20A,Glowacki-22,Manuwal-22}. We instead use the baryonic mass since it includes the gas mass, which is an important consideration for low-mass dwarf galaxies that are typically gas-dominated. Furthermore, it has been well established that a galaxy's baryonic mass and its total dynamical mass are closely correlated, as demonstrated by the baryonic Tully-Fisher relation \citep[e.g.][]{McGaugh-00,Verheijen-01-4,Ponomareva-21,Gogate-23}.

Fig.~\ref{fig:Asymm-MbarFS} shows $\Aflux$ (top row) and $\Amod$ (bottom) as a function of baryonic mass, adopting the same marker and colour scheme used in Fig.~\ref{fig:AmodAfluxPlot} for the three categories. To tease out any trends that may be hidden by the large scatter in the left panels, the right panels display the median asymmetries of each category in bins of baryonic mass, with varying bin widths to ensure that the total number of galaxies per bin remains approximately constant. The vertical error bars and shaded regions depict the standard error on the median and the interquartile range, respectively. 

From the left panels and our sample's baryonic mass distribution (top-right panel of Fig.~\ref{fig:FS_massDist}), it is evident that the sample suffers from incompleteness at the low-mass end where there are only galaxies from the \mbox{\simba-25} sample. We fit a single Schechter function \citep{Schechter-76} to \mbox{\simba-25's} baryonic mass function (BMF, see Appendix~\ref{sec:appendixSchechter}) and find that the sample is complete down to $M_{\text{baryon}}\approx 10^{9.5}~\text{M}_{\odot}$, before deviating ($\geq 0.25$~dex) from the best-fitting Schechter function. Since it is unclear how incompleteness at the low-mass end may impact the reliability of any observed trends between \HI asymmetry and baryonic mass in this mass regime, only galaxies with $M_{\text{baryon}} \geq 10^{9.5}~\text{M}_{\odot}$ are considered when plotting the median trends. We discuss possible reasons for the apparent absence of highly asymmetric galaxies in the low baryonic mass regime in Section~\ref{sec:lowAsymmDiscussion}.

As can be seen from the top right panel of Fig.~\ref{fig:Asymm-MbarFS}, merged galaxies have the highest median $\Aflux$ values in all mass bins, while interacted galaxies have comparable or marginally higher median $\Aflux$ values than isolated galaxies, supporting the results from Section~\ref{sec:AsymmDist}. However, we find no trend between $\Aflux$ and baryonic mass for any category. In contrast, the bottom right panel of Fig.~\ref{fig:Asymm-MbarFS} shows a clear inverse trend of decreasing $\Amod$ with increasing baryonic mass for the interacted and merged categories. There are hints of an inverse correlation between $\Amod$ and baryonic mass for the isolated category, but the trend is much flatter than the trends observed for the interacted and merged categories. These results suggest that larger galaxies with deeper potential wells, and thus greater gravitational restoring forces, tend to be relatively more symmetric, while the \HI discs of lower mass galaxies are more susceptible to being distorted by galaxy-galaxy interactions and mergers. We discuss the differences in trends for $\Aflux$ and $\Amod$ and the implications for upcoming \HI surveys in Section~\ref{sec:D_Implications}.

\section{Discussion}\label{sec:Discussion}

In the following sections we place the findings presented in this work into the broader context of the existing literature, provide possible explanations for unusual results, and discuss implications for future work.

\subsection{Comparison with previous work}

\subsubsection{The \HI size-mass relation}\label{sec:DiscussHI_MD}

As outlined in Section~\ref{sec:DataVerification}, we find that the \HI size-mass relations for our \mbox{\simba-25} and \mbox{\simba-100} samples broadly match the observed relation from \cite{Wang-16}, but galaxies with $M_{\text{\textsc{H\,i}}} \gtrsim 10^{9.5}~\text{M}_{\odot}$ tend to fall below the relation. Despite masking the data cubes, we are still sensitive to \HI emission down to $\sim \! 3\times 10^{18}$~cm$^{-2}$ (i.e. an order of magnitude lower than the $3\sigma$ column density sensitivity of most completed interferometric \HI surveys). Coupled with the fact that the data cubes are noise-free, this means that our \HI measurements include faint \HI emission that would typically be missed in real interferometric observations. Furthermore, the inclusion of faint \HI has a larger impact on the measured \HI mass than $R_{\text{\mbox{\textsc{H\,i}}}}$, since the latter is by definition measured at the radius where $\Sigma_{\text{\mbox{\textsc{H\,i}}}} = 1$~M$_{\odot}~$pc$^{-2}$, which corresponds to a column density of $\sim \! 1.249\times 10^{20}$~cm$^{-2}$.

More recently, \cite{Wang-24} have combined single dish data from the Five-hundred-metre Aperture Spherical radio Telescope and the Green Bank Telescope with archival interferometric data from the Very Large Array to recover missing \HI flux from the latter due to the lack of short baseline spacings. With updated \HI measurements for 13 galaxies, they find that the galaxies shift rightwards by varying amounts in the \HI size-mass plane (see their fig.~8), resulting in a slightly shallower relation and more galaxies that lie $\geq 1\sigma$ below the previously established relation from \cite{Wang-16}. While a larger sample is needed to determine whether the \HI size-mass relation is indeed shallower, these results tentatively suggest that \simba's relation agrees more with observations that include the diffuse \HI component. 

\subsubsection{The relative fractions of asymmetric galaxies}\label{sec:Discuss_AfluxRate}

As already mentioned, various studies have compared the prevalence of asymmetric global \HI profiles in different galaxy populations in an effort to infer the probable external mechanisms responsible for the observed asymmetries. Our result of a higher fraction of galaxies with $\Aflux > 0.12$ in the merged category relative to the isolated category (46~per~cent vs. 28~per~cent, see Table~\ref{table:AfluxRates}) qualitatively supports the findings of \cite{Bok-19} but is in conflict with \cite{Zuo-22}, who found no significant difference in the levels of spectral asymmetry between their merger and control samples. It should be noted that \cite{Zuo-22} use the curve of growth approach introduced in \cite{Yu-20}, as well as different velocity limits, to quantify spectral asymmetries. Furthermore, the experimental samples of both studies consist of galaxies in different merging stages and their control samples have different selection criteria, making a direct comparison between our results and theirs difficult. 

Another point of tension between our results and the literature is the relatively high fraction of asymmetric global \HI profiles in our isolated category. For example, \cite{Espada-11} found that only 2~per~cent of their isolated AMIGA sample exceeded their 3$\sigma$ threshold, whereas 15~per~cent of galaxies in our isolated category are found to have $\Aflux > 0.16$. \cite{Espada-11} do not mention the mass range of their sample, but the lack of a correlation between $\Aflux$ and baryonic mass presented in this work suggests that this is not the main reason for any discrepancy between our results and theirs. However, one possible reason could again be the absence of noise in our data cubes. While it has been shown that noise will artificially inflate spectral asymmetries in low signal-to-noise data \citep[e.g.][]{Watts-20A,Deg-20,Bilimogga-22}, it could also lead to a decrease in the measured asymmetry if one half of a very lopsided global \HI profile has flux density values that are comparable to the noise level of an observation. Similarly to what was noted by \citet[][see their figs.~10 and 11]{Watts-20B}, we find a number of asymmetric global \HI profiles in the isolated category where one half of the profile consists entirely of faint emission. An example of this is shown in the bottom right panel of the third row of Fig.~\ref{fig:AmodAfluxPlot}. If a given observation is not sensitive enough to detect the low flux density half of such a global profile, this could lead to an underestimation of the intrinsic spectral asymmetry as the profile would likely be mischaracterised as a single-peaked profile. It is also worth noting that the AMIGA data were taken with large single dish telescopes with relatively small primary beams, thereby suppressing the sensitivity to any asymmetries present in the \HI disc outskirts of the targeted galaxies.

With respect to \HI morphological asymmetries, spatially asymmetric, diffuse, low column density \HI could also exist below the column density sensitivity of interferometric observations, leading to a lower rate of morphological asymmetries measured in observational data. The presence of this diffuse \HI in our simulated isolated galaxies could be due to the accretion of cold gas from the cosmic web or feedback-driven outflows, as \cite{Manuwal-22} found that EAGLE galaxies with asymmetric global profiles are subject to stronger outflows and more elevated accretion rates than those with symmetric global profiles. On the other hand, it is possible that \simba\ and other cosmological simulations may overproduce the amount of diffuse \HI in galaxies, and that this results in more asymmetric \HI distributions since the diffuse \HI is more easily disturbed by internal and external mechanisms. However, deeper \HI observations for a statistical sample are crucial to determine if the excess diffuse \HI predicted by simulations is in fact present but hidden in the noise of current observations, or if there is further tension between simulations and observations \citep[see e.g.][]{Marasco-25}.

\subsubsection{The correlation between \HI spectral and morphological asymmetry}\label{sec:Discuss_AsymmCorr}

Our result of a weak correlation is broadly in agreement with the findings of \cite{Reynolds-20A} but is in conflict with those of \cite{Bilimogga-22}, who found no correlation between $\Aflux$ and $\Amod$ for their sample of EAGLE galaxies. Since our mock samples were generated in similar manners and we adopt the same rotation centre and column density threshold used in \cite{Bilimogga-22} when calculating $\Amod$, it is unclear why our results disagree. One possible reason for this tension could be that we measure $\Amod$ at the resolution at which $N_{\text{beams}}\approx10$, whereas they measure $\Amod$ at a fixed spatial resolution of $56\arcsec$ (corresponding to a physical scale of $\sim$4.62~kpc at the distance of 17~Mpc). To test this, we recalculate the Spearman rank test using the $\Amod$ values calculated at a fixed spatial resolution of $15\arcsec$, which corresponds to a physical scale of $\sim$5.22~kpc, and obtain a marginally weaker correlation coefficient of $r_{s} = 0.34$ ($p$-value $< 0.001$). Thus, our resolution approach in this work has a slight impact on the correlation coefficient measured, but is likely not the reason for the discrepancy between our results and those of \cite{Bilimogga-22}. Another reason could be that the differences in how various feedback mechanisms are implemented in \simba\ and EAGLE lead to overall different spatial distributions of \HI in the simulations. However, a full investigation into the cause of this conflict is beyond the scope of this study.

\subsection{On the absence of asymmetric galaxies at low baryonic masses} \label{sec:lowAsymmDiscussion}

While investigating the relationship between \HI asymmetry and baryonic mass in Section~\ref{sec:Mbar-HIasymm}, we noted the apparent lack of highly asymmetric galaxies below the sample's mass completeness limit of $M_{\text{baryon}} \approx 10^{9.5}~\text{M}_{\odot}$. This is somewhat counter-intuitive since low-mass galaxies are expected to host more asymmetric \HI discs due to their shallower potential wells. A number of studies have observed that highly asymmetric galaxies tend to have lower \HI gas fractions compared to their symmetric counterparts \citep[e.g.][]{Watts-20A,Glowacki-22}. With this in mind, we checked whether low-mass galaxies below the completeness limit are preferentially more \HI\!-rich than the rest of the sample as a possible explanation and found that this is not the case.

We suspect that the absence of high \HI asymmetries at the low-mass end is largely a consequence of the \HI mass cuts imposed during our sample selection. As mentioned in Section~\ref{sec:sample}, we applied the same \HI mass cuts to both the expected \HI mass provided by \caesar\ and the \HI mass that is confined to the star-forming region of the galaxy, to ensure that the inner parts of the \HI discs were sufficiently well-resolved. Thus, it seems plausible that galaxies with \HI masses close to the \HI mass threshold may have truncated, relatively symmetric \HI discs with the vast majority of the \HI centrally concentrated within the stellar disc. Additionally, the incompleteness of the sample with respect to the BMF, which likely impacts the sampling of galaxies from the different categories, also hinders our ability to make any firm conclusions on the trends observed below the completeness limit.

\subsection{Implications for future studies}\label{sec:D_Implications}

Throughout Section~\ref{sec:DriversofAsymm}, we observed a more distinct separation between the distributions of $\Amod$ for galaxies with different dynamical histories compared to $\Aflux$, which systematically underestimated the degree of asymmetry in the galaxies' \HI discs. As previously stated, the shape of a galaxy's global \HI profile is a convolution of the galaxy's kinematics and the spatial distribution of its \HI disc, meaning that the profile's shape (and thus the measured $\Aflux$ value) can vary significantly depending on the galaxy's inclination and how an asymmetric feature is oriented relative to the line-of-sight \citep[e.g.][]{Manuwal-22,Deg-23}. It is also worth noting that $\Aflux$ will not adequately characterise asymmetries in a global profile if said asymmetries, such as variations in the slopes of the profile edges, do not result in differences in the areas of the two profile halves. Thus, in the context of ongoing and future untargeted \HI surveys where the majority of direct detections will be spatially unresolved, these results suggest that $\Aflux$ should be considered a lower limit on the intrinsic \HI asymmetry, provided that the data used are of sufficient quality in terms of their signal-to-noise and spectral resolution \citep[see e.g.][]{Watts-20A,Deg-20}. In particular, it must be emphasised that a low $\Aflux$ value cannot be used to infer that an \HI disc is undisturbed. 

Regarding future studies with access to large samples of spatially resolved \HI observations, we have shown that failing to account for differences in spatial resolution can introduce unexpected biases. To fairly compare galaxies within a survey spanning a wide range of \HI masses, it is important to match the relative spatial resolution across the sample when calculating morphological asymmetries. Alternatively, another approach would be to group galaxies into subcategories based on their spatial resolution to avoid reducing the resolution of the entire sample to that of the least spatially resolved galaxy. However, with the second approach there is the risk of further dividing what may already be a limited sample. We have also demonstrated that baryonic mass is another crucial factor to consider when comparing the \HI morphological asymmetries obtained from datasets across different surveys. If one wishes to investigate the trends between \HI asymmetries and environment, care must be taken to match samples in terms of their masses.

Lastly, it is important to acknowledge that while $\Amod$ is generally better at characterising disturbed \HI morphologies, the overlap of different categories at various $\Amod$ values indicates that $\Amod$ on its own is insufficient to deduce the most probable mechanism responsible for the observed \HI asymmetry of a galaxy. As such, applying a threshold in $\Amod$ would not result in a clear separation between our galaxy categories. However, incorporating an analysis of the \HI kinematics could assist in the interpretation of \HI asymmetries. For example, \cite{Serra-24} recently presented a detailed investigation of dwarf galaxy NGC~1427A in the Fornax cluster, in which they convincingly argue that the complex \HI morphology and kinematics of the system results from ram pressure being exerted by the intracluster medium in the aftermath of a tidal interaction (or merger) between two galaxies. Such an analysis is beyond the scope of this study, but it is a compelling direction worth pursuing in our future work with the MDS. 

\section{Conclusions}\label{sec:Conclusions}

We have used the \simba\ cosmological simulations to investigate the relationship between the recent dynamical histories 
and \HI spectral and morphological asymmetries of $\sim$1100 spatially resolved galaxies at low redshift. We created synthetic \HI data cubes designed to match \HI observations from the MDS to enable the extraction of \HI data products in an observational manner, and quantified asymmetries in the global \HI profiles and moment-0 maps using the flux ratio asymmetry ($\Aflux$) and the modified asymmetry parameter ($\Amod$), respectively. We placed galaxies into three main categories (isolated, interacted, or merged) based on their dynamical histories over the preceding $\sim$2~Gyr, and defined a refined isolated category to check the robustness of our results for the isolated category. We then compared the \HI asymmetry distributions of the various categories and explored trends between asymmetry and baryonic mass. Our findings are summarised as follows:

\begin{enumerate}
    \item Galaxies from \simba\ follow the observed \HI size-mass relation from \cite{Wang-16} relatively well, but tend to have slightly smaller \HI diameters at $M_{\text{\textsc{H\,i}}} \gtrsim 10^{9.5}~\text{M}_{\odot}$. This likely owes to the absence of noise in the mock data cubes, which can have a larger impact on the measured \HI mass than the \HI diameter when more diffuse \HI is recovered in the disc outskirts \citep[see e.g.][]{Wang-24}.
    \item A spurious trend between \HI mass and $\Amod$ is introduced when measuring $\Amod$ at a fixed resolution, which is largely a secondary effect of the \HI size-mass relation. This trend disappears when $\Amod$ is measured using the moment-0 map smoothed to the spatial resolution at which a galaxy is resolved by approximately 10 beams in its area. This highlights that the relative spatial resolution must be taken into account when comparing $\Amod$ values in a sample that spans a wide range of \HI masses.
    \item The cumulative distributions for both $\Aflux$ and $\Amod$ show that the refined isolated and isolated categories are statistically indistinguishable from one another, indicating that our results are insensitive to the chosen isolation selection criteria. 
    \item Galaxies in the interacted category are, on average, more asymmetric than galaxies in the isolated category, but a systematic difference between the two categories' asymmetry distributions is only found with $\Amod$ and not $\Aflux$. Galaxies in the merged category are generally even more asymmetric, with statistically different asymmetry distributions compared to both the isolated and interacted categories. These results suggest that galaxy mergers and interactions are likely responsible for the higher mean asymmetries observed in these categories. 
    \item For the interacted and merged categories, an inverse correlation exists between baryonic mass and $\Amod$, suggesting that massive galaxies are typically more symmetric, whereas lower mass galaxies are more prone to having their \HI discs distorted from interactions and mergers. This trend is not seen between baryonic mass and $\Aflux$ and is likely due to $\Aflux$ underestimating the degree of asymmetry of a galaxy's \HI disc compared to $\Amod$. 
\end{enumerate}

Overall, these results showcase the importance of spatially resolved \HI observations and the need for caution when relying solely on the global \HI profiles of galaxies to study \HI asymmetries in relation to other galaxy properties. Looking to the future, a natural progression of this work will be to incorporate kinematic asymmetries and expand our analysis to observations from the MDS. This is a rich dataset with \HI detections across a range of cosmic environments, from voids to filaments and clusters, allowing us to investigate the average gas content of various galaxy populations and how different environmental mechanisms impact the \HI reservoirs of galaxies.

\section*{Acknowledgements}

The authors thank the anonymous referee for constructive comments and suggestions. We acknowledge helpful discussions with the ASymba team, N. Hatamkhani, J. Healy and R. Talens. We also thank J. Wang for providing the observational data used in Fig.~\ref{fig:medRadProf}.

NH acknowledges support from the National Research Foundation of South Africa (Grant number: 120224) and the Leids Kerkhoven–Bosscha Fonds (LKBF). KAO acknowledges support by the Royal Society through Dorothy Hodgkin Fellowship DHF/R1/231105. MG is supported by the Australian Government through the Australian Research Council's Discovery Projects funding scheme (DP210102103), and through UK STFC Grant ST/Y001117/1. MG acknowledges support from the Inter-University Institute for Data Intensive Astronomy (IDIA). IDIA is a partnership of the University of Cape Town (UCT), the University of Pretoria and the University of the Western Cape.

We acknowledge the use of the ilifu cloud computing facility (\url{www.ilifu.ac.za}), a partnership between UCT, the University of the Western Cape, the University of Stellenbosch, Sol Plaatje University, the Cape Peninsula University of Technology and the South African Radio Astronomy Observatory. The ilifu facility is supported by contributions from IDIA, the Computational Biology division at UCT and the Data Intensive Research Initiative of South Africa (DIRISA). This work has made use of the following Python packages: Astropy \citep{Astropy-18}, Matplotlib \citep{Matplotlib-07}, NumPy \citep{Numpy-11}, and SciPy \citep{Scipy-20}. This work has also made use of NASA's Astrophysics Data System. For the purpose of open access, the author has applied a Creative Commons Attribution (CC BY) licence to any Author Accepted Manuscript version arising from this submission.

\section*{Data Availability}

The \simba\ simulation snapshots and accompanying \caesar\ catalogues are publicly available at \url{http://simba.roe.ac.uk}. The data underlying this work will be made available upon reasonable request to the corresponding author.

\bibliographystyle{mnras}
\bibliography{References} 

\appendix

\section{Moment-0 Maps and global H I profiles}\label{sec:AppendixMom0}

\begin{figure*}
 \centering\includegraphics[width=\linewidth]{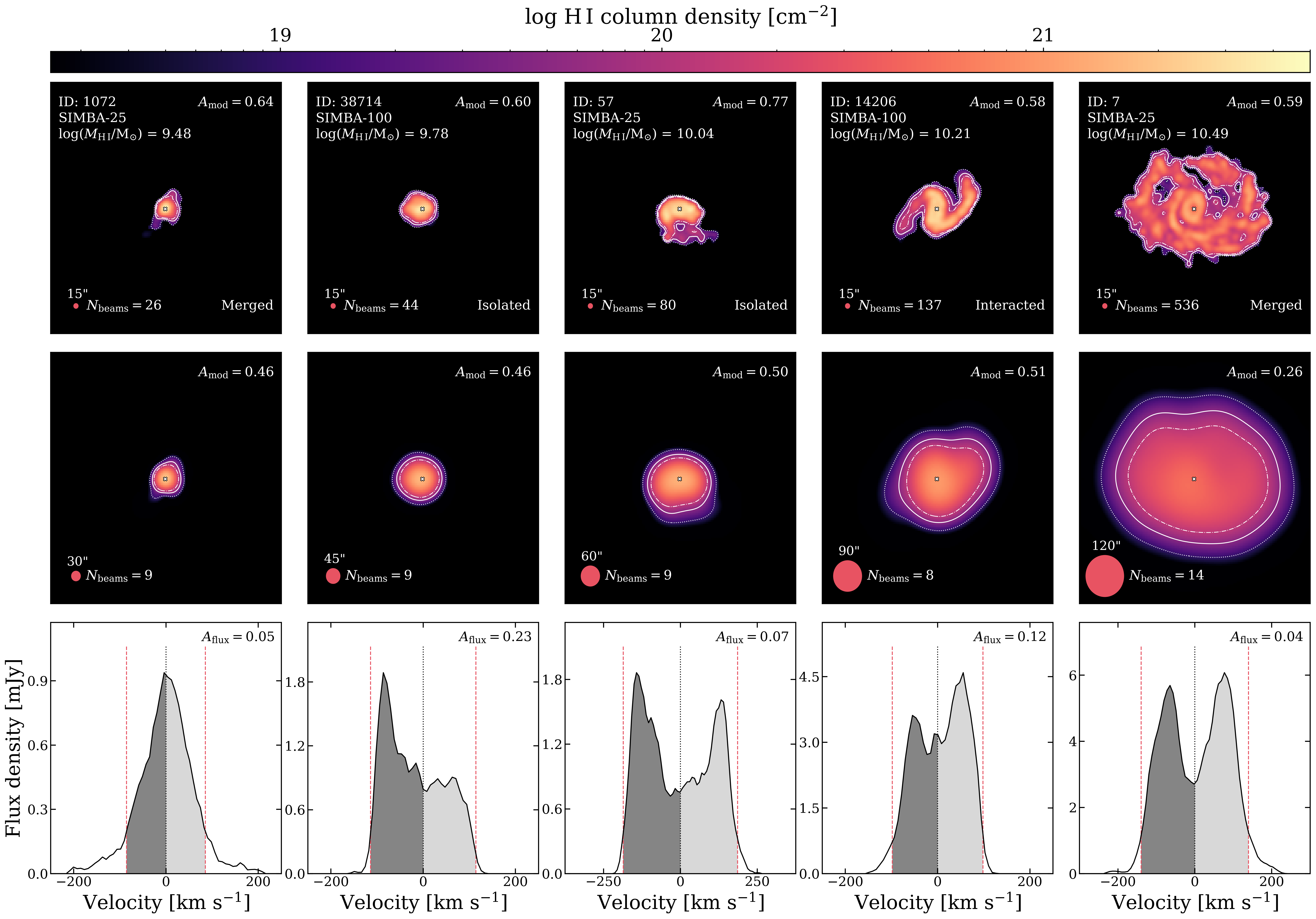}
 \caption{Moment-0 maps (top and middle rows) and global \HI profiles (bottom row) of five galaxies in our sample. See the text of Appendix~\ref{sec:AppendixMom0} for more details.}
 \label{fig:Mom0-HIProfiles}
\end{figure*}

Fig.~\ref{fig:Mom0-HIProfiles} shows the moment-0 maps (top and middle rows) and global \HI profiles (bottom row) of five galaxies, arranged from left to right in order of increasing \HI mass. For each galaxy, the moment-0 maps are shown at $15\arcsec$ resolution (top panel) and the resolution at which $N_{\text{beams}}\approx10$ (middle). The value of $N_{\text{beams}}$ is given in the bottom left corner of the panels in the top and middle rows next to the ellipse showing the size of the beam. Each panel presenting the moment-0 maps is 250~kpc across and centred on the minimum of the galaxy's gravitational potential, which is shown by a white cross. The white dash-dotted, solid, and dotted contours are at column densities of $1,5, \text{ and }12.5\times 10^{19}$~cm$^{-2}$. For each global \HI profile, the dashed pink lines represent $v_{l,20}$ and $v_{h,20}$, and the black dotted line represents $v_{\text{sys}}$. The galaxy ID, volume, logarithm of the \HI mass and the category are given in the top panels. The measured $\Amod$ and $\Aflux$ values are quoted in the top right corner of the relevant panel.

\section{Determining the incompleteness at low baryonic masses}\label{sec:appendixSchechter}

One way to assess the incompleteness of a sample is to fit a model function to the data and establish where the fit deviates significantly. Numerous studies \citep[see e.g.][]{Salucci-99,Bell-03,Papastergis-12,Eckert-16} have shown that the BMF is well described by a single Schechter function, given by

\begin{equation}
    \phi(M_{\text{baryon}}) = \ln(10) \phi_{\ast} \bigg( \dfrac{M_{\text{baryon}}}{M^{\ast}} \bigg)^{(\alpha+1)} \exp \biggl( \dfrac{-M_{\text{baryon}}}{M^{\ast}} \biggr), \label{eq:Schechter}
\end{equation}

where $M^{\ast}$ is the characteristic `knee' mass above which the number of galaxies declines exponentially, $\alpha$ sets the slope of the power-law at the low-mass end, and $\phi_{\ast}$ is the normalisation constant. 

We use the \mbox{\simba-25} sample to determine the completeness limit since the incompleteness is most apparent at the low-mass end of the baryonic mass distribution, where there are only galaxies from this sample. Additionally, attempting to fit the entire sample introduces complications because the two samples are taken from different volumes. 

\begin{figure}
	\includegraphics[width=\columnwidth]{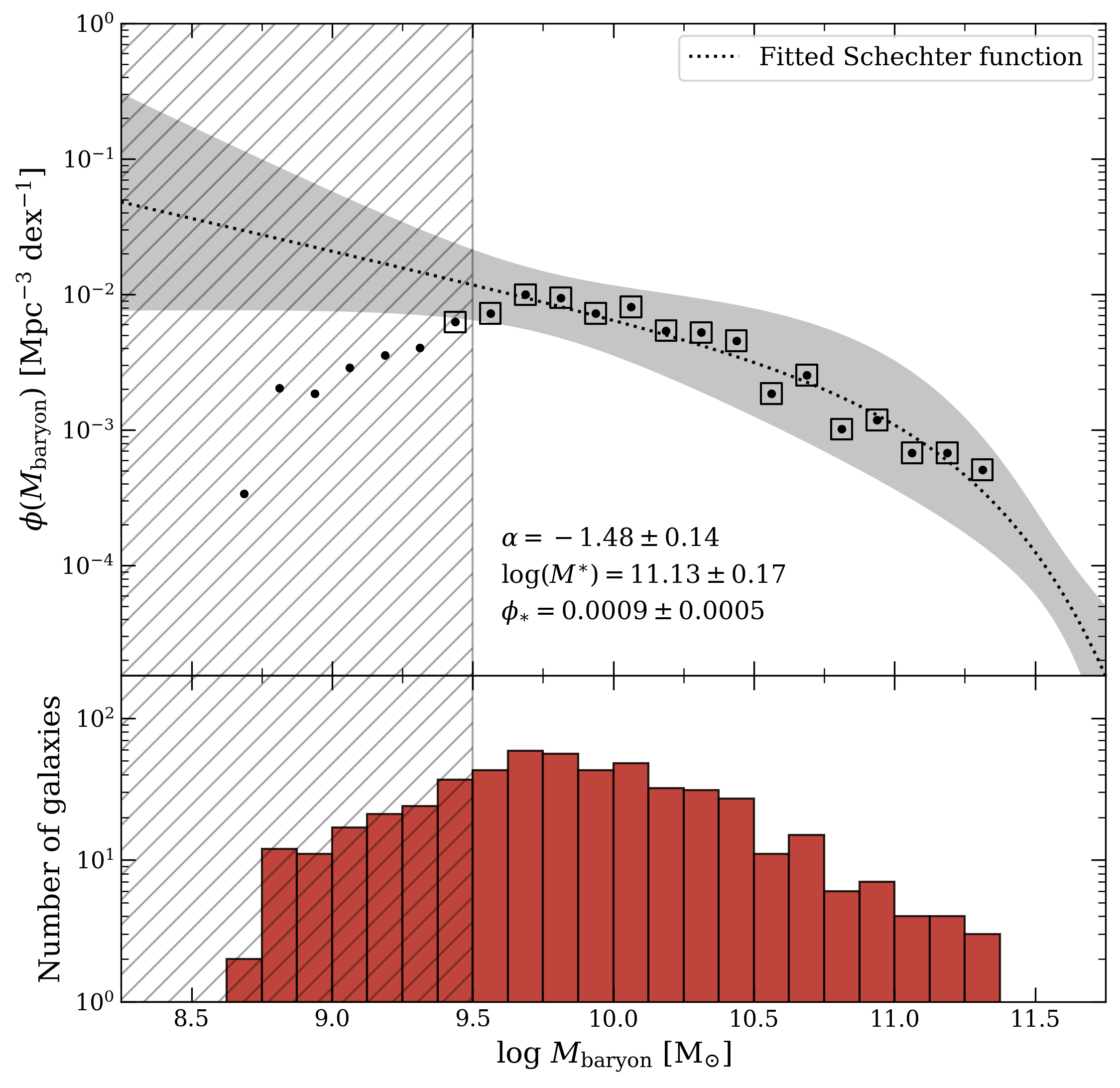}
    \caption{The baryonic mass function of the \mbox{\simba-25} sample (black circle markers). The open square symbols represent the data points used to fit the single Schechter function. The parameters of the best fit, which is shown by the black dotted line, are quoted in the top panel. The shaded grey region shows the 3$\sigma$ uncertainty of the fit. The histogram of the baryonic mass distribution is shown in the bottom panel. The hatched region indicates where the sample is incomplete.}
    \label{fig:SchechterFit}
\end{figure}

A nonlinear least-squares fitting routine is used to fit Eq.~(\ref{eq:Schechter}) to the mid- to high-mass end of \mbox{\simba-25's} BMF, and the uncertainties are computed from the covariance matrix. The best-fitting Schechter function and the 3$\sigma$ uncertainty of the fit are shown in Fig.~\ref{fig:SchechterFit} as the black dotted line and grey shaded region, respectively. For transparency, we include the parameters of the best fit in the figure, but stress that we are not attempting to compare the results of the fit to those found in the literature. We estimate the baryonic mass completeness limit to be $M_{\text{baryon}} = 10^{9.5}~\text{M}_{\odot}$ as the BMF and the best Schechter fit are in broad agreement down to this mass, before the data diverges by more than 0.25~dex from the fit (shown by the hatched region in the figure).

\bsp	
\label{lastpage}
\end{document}